\documentclass{article}
\usepackage{graphicx} 
\usepackage{latexsym}
\usepackage{amssymb}
\usepackage{amsthm}
\usepackage{amsmath}
\usepackage{verbatim}
\usepackage{fancyvrb}
\usepackage{algorithmic}
\usepackage{algorithm}

\newcommand{\fa}{\forall\hspace{0.5ex}}
\newcommand{\ex}{\exists\hspace{0.5ex}}

\newcommand{\done}{\hspace{1em}$\Box$}

\newcommand{\bfit}[1]{\textbf{\textit{#1}}\index{#1}} 

\newcommand{\ddd}{\cdots}

\newcommand{\combin}[2]{{\ensuremath
\left(\begin{array}{c}#1\\#2\end{array}\right)}}
\newcommand{\blankline}{\ \\\indent}

\renewcommand{\emph}[1]{\textbf{#1}}
\renewcommand{\bfit}[1]{\textbf{\textit{#1}}}

\newlength{\theoskip}
\newlength{\theoskipAlgo}
\newtheorem{lemm1}{Lemma}[section]
\newtheorem{rule1}{Rule}
\newtheorem{prop1}{Property}[section]
\newtheorem{propos1}{Proposition}[section]
\newtheorem{paradox1}{Paradox}[section]
\newtheorem{theo1}{Theorem}[section]
\newtheorem{coro1}{Corollary}[section]
\newtheorem{proposi1}{Proposition}[section]
\newtheorem{obsv1}{Observation}[section]
\theoremstyle{remark}  
\newtheorem{remk1}{Remark}[section]
\theoremstyle{definition} 
\newtheorem{exam1}{Example}[section]
\newtheorem{defi1}{Definition}[section]
\newtheorem{algo1}{Algorithm}[section]
\newtheorem{func1}{Function}[section]
\newtheorem{query1}{Query}[section]
\newtheorem{expl1}{Pf\hspace*{-.8ex}}[section]
\newtheorem{proc1}{Procedure}[section]

\newenvironment{lemm}{\vspace{\theoskip}\begin{lemm1}}{\end{lemm1}}

\newenvironment{obsv}{\vspace{\theoskip}\begin{obsv1}}{\end{obsv1}}

\newenvironment{func}{\vspace{\theoskip}\begin{func1}}{\end{func1}}

\newenvironment{defi}{\vspace{\theoskip}\begin{defi1}}{\end{defi1}}
\newenvironment{exam}{\vspace{\theoskip}\begin{exam1}}{\end{exam1}}

\newenvironment{proc}{\vspace{\theoskip}\begin{proc1}}{\end{proc1}}

\newcounter{mylisti}
\newcounter{mylista}
\newcounter{mylist1}

\newenvironment{lista}{\begin{list}{(\alph{mylista})}
   {\usecounter{mylista} \itemsep=0ex \topsep=0mm
            \parsep=0ex \parskip=0ex \partopsep=0mm }}
   {\end{list}}
\newenvironment{list1}{\begin{list}{(\arabic{mylist1})}
   {\usecounter{mylist1} \itemsep=0ex \topsep=0mm
             \parsep=0ex \parskip=0ex \partopsep=0mm}}
   {\end{list}}
\newenvironment{tab}{\begin{list}{}
  { \itemsep=0ex \topsep=0mm \parsep=0ex \parskip=0ex \partopsep=0mm} \item}
   {\end{list}}

\usepackage{algorithmic}
\usepackage{algorithm}

\newcommand{\angles}[1]{\langle#1\rangle}
\newcommand*\aw[1]{\doaw#1\relax}
\def\doaw#1,#2\relax{{#1\angles{#2}}}
\newcommand{\rar}{\rightarrow}

\newcommand{\dd}{\hspace*{-.6ex}\cdot\hspace*{-.6ex}\cdot}
\newcommand{\calF}{{\cal F}}

\title{Discovery of Approximate Differential Dependencies}
\author{Jixue Liu $^a$ \hspace{1ex} Selasi Kwashie $^a$\hspace{1em}
Jiuyong Li $^a$\hspace{1em} \\
Feiyue Ye $^b$\hspace{1em} Millist Vincent $^a$\\
 $^a$ University of South Australia, \hspace{1em} \\
 $^b$ Jiangsu University of Technology \\
 \{jixue.liu, selasi.kwashie, jiuyong.li\}@unisa.edu.au\\
 millist.vincent@unisa.edu.au, \ \ yfy@jstu.edu.cn
}

\begin{document}

\date{}\maketitle
\begin{abstract}
Differential dependencies (DDs) capture the relationships between
data columns of relations. They are more general than functional
dependencies (FDs) and and the difference is that DDs are defined on
the distances between values of two tuples, not directly on the
values. Because of this difference, the algorithms for discovering
FDs from data find only special DDs, not all DDs and therefore are
not applicable to DD discovery. In this paper, we propose an
algorithm to discover DDs from data following the way of fixing the
left hand side of a candidate DD to determine the right hand side.
We also show some properties of DDs and conduct a comprehensive
analysis on how sampling affects the DDs discovered from data.
\end{abstract}
{\bf keywords}: {Differential dependencies, functional dependencies,
dependency discovery, knowledge discovery, lattice, partition}

\section{Introduction}

Data quality has been a core concern in data management. Many types
of constraints, like logic constraints, keys, referential
constrains, and functional dependencies, have been designed to
control data quality in databases. Differential dependency
\cite{diffDep-tods11} is a new type of such dependencies. It also
generalizes metric functional dependencies defined in
\cite{metricFD-icde09koudas}.

Differential dependencies (DDs) are defined to constrain the
closeness of the values of dependent attributes with respect to the
closeness of the values of determinant attributes. More specifically,
the differential dependency $\aw{X,w}\rightarrow A\angles{w_1}$
requires that when the distance of any two tuples on the attribute
set $X$ is within the range $w$, the distance of the same two tuples
on the attribute $A$ are within the range $w_1$. For example,
$\aw{Age,[0,0]}\rightarrow \aw{Sal,[0,1]}$ is a DD that requires that
if the $Age$ difference of two tuples is 0, their $Sal$ difference
should be no more than $1$. This DD is satisfied by the data in Table
\ref{tabl-motivat}. $\aw{Age,[0,0]}$ is called the left hand side
(lhs) and $\aw{Sal,[0,1]}$ the right hand side (rhs).

\begin{table}[h]
\center \caption{ A table satisfying $\aw{Age,[0,0]}\rar
\aw{Sal,[0,1]}$ but violating $\aw{Age,[0,0]}\rar \aw{Sal,[0,0]}$}
\label {tabl-motivat}
\begin{tabular}{|l|l|l|l|l|}\hline 
 tid    & Age    &  Edu  & Sex &  Sal \\ \hline
 $t_1$  & 20     &    3  &  0  & 3  \\ \hline
 $t_2$  & 20     &    3  &  1  & 3  \\ \hline
 $t_3$  & 20     &    4  &  0  & 4  \\ \hline
 $t_4$  & 25     &    5  &  1  & 5  \\ \hline
\end{tabular}\\
{\small $Edu$ means education completed: 3=Bachelors  4=Masters
5=PhD}
\end{table}

Differential dependencies are defined on closeness of tuples and the
closeness is relative. They do not rely on absolute values. In other
words, to satisfy $\aw{Age,[0,0]}$, the $Age$ values of two tuples
can be $20$ and $20$, and can also be $50$ and $50$; in both cases,
the difference is $0$.

Like functional dependencies and matching dependencies
\cite{md-pvldb09}, differential dependencies can be used in many
applications. They can be used to warrant data quality like what
function dependencies do. For example, a DD can be defined to
require that if two post codes of two locations are close, their
addresses must be in the same city or the distance between their
addresses must be within a certain range.

Differential dependencies can also be used to detect data quality
issues in data cleaning. If a discovered DD shows that two
attributes that should take close values have taken values that are
far apart, their data has quality problems. This gives guide to data
cleaning processes to quickly identify problems.

Differential dependencies holding on data may describe patterns that
are new to domain experts. They enrich the knowledge in the
application areas \cite{similar-rela-dasfaa06belo}. In addition, the
knowledge represented by the DDs can be used in query processing and
data management as it is done in inductional databases
\cite{inductiveDB-sigkddexpl02raedt}.

Among many of these applications, especially in knowledge discovery
and data cleaning, the discovery of DDs from existing data is a
critical task. Because DDs are more general than other types of
dependencies like FDs and match dependencies etc, some of the
patterns that are interesting and can be described in DDs cannot be
described in previous dependencies. This raises two problems. One is
that the algorithms used to find previous dependencies will not find
DD specific patterns. The other is that if we want to know these DD
specific patterns, new algorithms much be developed. This motivates
the work in this paper and developing an effective and efficient
algorithm is the objective of this paper.

The most relevant work on DD discovery to our work is the reduction
algorithm proposed in \cite{diffDep-tods11}. It uses a
fix-rhs-reduce-lhs approach and reduces the search space containing
subsumption-ordered nodes to find DDs for each rhs attribute and
every of its interval. This approach involves storing an
exponentially sized search space, which has a severe performance
bottleneck and  makes the applicability of the algorithm very much
limited to relations with very small number of attributes. At the
same time, the distance intervals of attributes are assumed to start
from 0 in its experiments. This leads to some possible DDs not to be
found. The detailed analysis of these is given in the section of
Related Work of this paper. In contrast, our algorithm follows a
fix-left-find-right approach: it checks all possible lhs candidates
and determines the rhs and it finds all DDs.

In this paper, we show that DDs have a property that given a lhs for
a DD, a rhs can always be found if its interval is large enough.
Thus, there is a possibility that the rhs interval is affected by
outliers in the data which reduces the usefulness of discovered
dependencies. We propose to use approximate satisfaction so that
outlier data points can be identified and ignored.

We propose an algorithm to discover differential dependencies and
the algorithm is a partition based approach. Pruning rules are used
to reduce the search space. At the same time, we use two parameters,
support to reduce the complexity of the computation, and
interestingness to find only 'good' DDs. The interestingness
parameter is design to avoid DDs with trivially large rhs intervals.

We conduct a comprehensive analysis on how sampling, while it reduces
the size of the computation, introduces errors in the discovered DDs
and possible ways to filter out the errors. Sampling causes `wanted'
DDs to be missed, `unwanted' DDs to be discovered, and trivial DDs to
appear like non-trivial DDs.

Our experiments show that (1) our approach is effective in
identifying problems in data when DDs of full satisfaction and of
approximate satisfaction are compared; (2) our algorithm is
efficient in space and time consumption; (3) our algorithm discovered
outliers in the Adult data. (4) Errors of sampling can be
significant, but some of the errors can be filtered.

The rest of the paper is organized in the following way. In Section
\ref{sec:dd-defi}, preliminary definitions are given. Section
\ref{sec:satisfy} presents some properties of DD satisfaction.
Section \ref{sec:disc} presents the discovery algorithm and the
pruning techniques. Section \ref{sec:sampling} details how sampling
is done. Section \ref{sec:experiments} presents the experiment
results. Section \ref{sec:related} discusses the works in the
literature related to this work. The final section concludes the paper and
discusses future work.

\section{Differential dependencies and properties}\label{sec:dd-defi}

We use $R$ to denote a relation schema and $r$ a relation instance of $R$. $dom(A)$
denotes the domain of attribute $A$. For a set of attributes $X=\{A_1,\ddd,A_k\}$ and a
single attribute $B$, $XB$ means $\{A_1,\ddd,A_k,B\}$. $|X|$ returns the number of
elements in $X$.

Let $d_{A}(v_1,v_2)$ be a function calculating the distance of the
two values, $v_1$ and $v_2$, of attribute $A$. It is assumed that
$d_{A}(v_1,v_2)=d_{A}(v_2,v_1)\ge 0$. The distance function can be
defined in editing distances of text values, differences of numeric
values, or in other ways. A \emph{differential function (DF)} of
attribute $A$ with regard to a distance interval $w=[d_1,d_2]$ is a
boolean function and is defined by $\aw{A,w}= d_1 \le d_{A}(v_1,v_2)
\le d_2$. The two functions $left(w)$ and $right(w)$ return $d_1$ and
$d_2$ respectively. If $d_1=d_2$, the notation of the interval is
simplified to $w=[d_1]$.

Let $w_a=[d_1,d_2]$ and $w_b=[d_3,d_4]$ be two intervals of an
attribute $A$. The \emph{order} $w_a \le w_b$ holds if $d_2 \le d_3$.
$w_a$ and $w_b$ are \emph{adjacent}, denoted by $w_a \leqq w_b$, if
$d_2=d_3$. The \emph{combination} of two adjacent intervals $w_a$ and
$w_b$, denoted by $w_a+ w_b$, is $[d_1, d_4]$.

A differential function of a set of attributes $X=\{A_1,\ddd,A_k\}$
with regard to a cube $w=w_1\times\ddd \times w_k =
[d_{11},d_{12}]\times \ddd \times [d_{k1},d_{k2}]$ is defined by
$\aw{X,w}=\aw{A_1,w_1}\ddd\aw{A_k,w_k}=\bigwedge_{j=1}^k
A_j\angles{w_j}$. Two cubes $w_a$ and $w_b$ on the same set of
attributes are \emph{adjacent} if their intervals on one of the
attributes are adjacent, and their intervals on all other attributes
are the same. The \emph{combination} of two adjacent cubes $w_a$ and
$w_b$, denoted by $w_a+ w_b$, is the smallest cube containing both
$w_a$ and $w_b$. In the rest of this paper, cubes are also referred
to as intervals.

Two differential functions $\aw{X,w_a}$ and $\aw{Y,w_b}$ are
\emph{joinable} if common attributes have the same interval in both
functions, i.e., for each $A\in (X\cap Y)$( $\aw{A,w} \in \aw{X,w_a}
\ and\ \aw{A,w} \in \aw{Y,w_b}$ ). The \emph{join} of two joinable
$\aw{X,w_a}$ and $\aw{Y,w_b}$, denoted by $\aw{X,w_a}\aw{Y,w_b}$, is
the cube containing all DFs of $\aw{X,w_a}$ and $\aw{Y,w_b}$, i.e.,
$\aw{A_1,w_1}\ddd\aw{A_n,w_n}$ where $n=|X\cup Y|$ and $\fa
\aw{A_i,w_i}[i\in[1,\ddd,n]](\aw{A_i,w_i}\in \aw{X,w_a} \ or\
\aw{A_i,w_i}\in \aw{Y,w_b})$.

A differential function $\aw{X,w_x}$ \emph{subsumes} a differential
function $\aw{Y,w_y}$, denoted by $\aw{X,w_x} \succeq \aw{Y,w_y}$, if
$\fa \aw{A_j,w_{xj}} \in \aw{X,w_x}$ ($\ex \aw{A_i,w'_{yi}} \in
\aw{Y,w_y} (w'_{yi}\subseteq w_{xj})$). That is, the subsuming
function (the left hand side) has less dimensions and larger
intervals.


\begin{defi}[Differential dependency (DD)]\cite{diffDep-tods11}
A DD is a formula $f=\aw{X, w_x} \rightarrow \aw{Y,w_y}$ where
$\aw{X, w_x}$ and $\aw{Y,w_y}$ are differential functions. A relation
$r$ satisfies $f$ if and only if for any two tuples $t_1$ and $t_2$
in $r$, if $\aw{X,w_x}$ returns true, $\aw{Y, w_y}$ returns true.
$\aw{X, w_x}$ is the lhs and $\aw{Y,w_y}$ is the rhs.
\end{defi}

%

\begin{defi}[Partition]
Given a set of attributes $X$ and an interval $w$ of $X$, the
\emph{tuple pair partition (partition, for short) for DF $\aw{X,w}$}
is a set of tuple pairs satisfying $\aw{X,w}$:
 \begin{align}\label{eq:F}
     F(\aw{X,w})=[\ (t_{p},t_{q})\ |\ (t_{p},t_{q}) \in r*r \land t_p\not=t_q\land  \fa A_i\angles{w_i} \in \aw{X,w} \ \ \\
(left(w_i)\le d_{A_i}(t_{p}[A_i],t_{q}[A_i])\le right(w_i)) \ ]
\notag
 \end{align}

Given a sequence of adjacent intervals $w_1,\ddd, w_k$, $w_i\le
w_{i+1}$ ($i=1,\ddd,k-1$) of the single attribute $B$, the
\emph{partition for attribute $B$} is defined to be the sequence of
labeled partitions for $B$'s intervals in order:
 \begin{equation}\label{eq:calF}
     \calF(B)=[w_1:F(\aw{B,w_1}),\ddd, w_k:F(\aw{B,w_k})]
 \end{equation}
 \done
\end{defi}

Tuple pair partitions for DFs are similar to the partitions for
relations. More details can be found in \cite{cosma85pods-parti}.

For performance reasons, tuple pairs in a partition are ascending
ordered. This order is defined to be $(t_{p},t_{q})<(t_{u},t_{v})$ if
$t_{p}< t_{u}$ or ($t_{p}= t_{u}$ and $t_{q}< t_{v}$). In this way,
operations on partitions can be done in linear time of their sizes.

\begin{exam}\label{exam-defi}
If we assume that $d_A(v_1,v_2)=abs(v_1-v_2)$ for any attribute $A$
in Table \ref{tabl-motivat}, the table satisfies $\aw{Age,[0]}\rar
\aw{Sal,[0, 1]}$ but violates $\aw{Age,[0]}\rar \aw{Sal,[0]}$. To see
these, we calculate partitions  and have

    $F(\aw{Age,[0]})=\{(t_1,t_2),(t_1,t_3),(t_2,t_3)\}$

    $F(\aw{Age,[5]})=\{(t_1,t_4),(t_2,t_4),(t_3,t_4)\}$

    $\calF(Age)=[0:\{(t_1,t_2),(t_1,t_3),(t_2,t_3)\},5:\{(t_1,t_4),(t_2,t_4),(t_3,t_4)\}]$

    $F(\aw{Sal,[0]})=\{(t_1,t_2)\}$

    $F(\aw{Sal,[1]})=\{(t_1,t_3),(t_2,t_3),(t_3,t_4)\}$

    $F(\aw{Sal,[2]})=\{(t_1,t_4),(t_2,t_4)\}$

    $F(\aw{Sal,[0, 1]})=\{(t_1,t_2),(t_1,t_3),(t_2,t_3),(t_3,t_4)\}$ \\
where, as examples, $F(\aw{Age,[0]})$ is the set of all tuple pairs whose $Age$ distances
are 0, and $F(\aw{Sal,[1]})$ the set of all tuple pairs whose $Sal$ distances are 1. We
notice that all the tuple pairs in $F(\aw{Age,[0]})$ are in $F(\aw{Sal,[0, 1]})$. So
$\aw{Age,[0]}\rar \aw{Sal,[0, 1]}$. Tested in the same way, $F(\aw{Age,[0]}) \not
\subseteq F(\aw{Sal,[0]})$, so $\aw{Age,[0]}\not\rar \aw{Sal,[0]}$. $\Box$
\end{exam}

Because the distance $d$ of two tuples on an attribute $A$ is a specific number, $d$
falls in only one of the two non-overlapping intervals $w_1$ and $w_2$ of $A$, but not in
both at the same time. Thus, we have the following properties for tuple pair partitions.

\begin{lemm}[Properties of partitions] \label{lemm:parti} \
 \begin{list1}
 \item $F(\aw{X,w_a}) \cap F(\aw{X,w_b}) = \phi$ if $w_a$ and $w_b$ do not overlap.
 \item $F(\aw{X,w_a+w_b}) = F(\aw{X,w_a})\cup F(\aw{X,w_b}) $ if $w_a$ and $w_b$ are adjacent.
 \item $F(\aw{X,w_a}\aw{Y,w_2})  =  F(\aw{X,w_a})\cap
 F(Y\angles{w_2})$ if $\aw{X,w_a}$ and $\aw{Y,w_2}$ are joinable.
 \end{list1}
\end{lemm}

Item (3) of the lemma tells how the partition for two joining DFs
should be calculated. This will be used in our algorithm later on.

In this paper, we are interested in DDs with single attributes on the
rhs like in the case of functional dependency discovery
\cite{huht99cj-tane-FD-AFD,liu10tkde-FDd}. The reason is that if we
know $\aw{X,w}\rar \aw{B,w_b}$ and $\aw{X,w}\rar \aw{C,w_c}$, we can
derive $\aw{X,w}\rar \aw{B,w_b}\aw{C, w_c}$ based on the inference
rules proposed in \cite{diffDep-tods11}.

With tuple pair partitions, the satisfaction of DD follows the lemma below.

\begin{lemm}\label{lemm:parti-satisfy} \
 \begin{lista}
 \item
  $F(\aw{X,w})\subseteq F(B\angles{w'})$ if and only if $\aw{X,w}\rightarrow
B\angles{w'}$.
  \item
  $F(\aw{X,w}B\angles{w'})= F(\aw{X,w})$ if and only if $\aw{X,w}\rightarrow
B\angles{w'}$.
  \item
  $|F(\aw{X,w}B\angles{w'})|=|F(\aw{X,w})|$ if and only if $\aw{X,w}\rightarrow
B\angles{w'}$.
 \end{lista}
\end{lemm}

The lemma is correct. (a) follows the DD definition. (b) is correct
because of Lemma \ref{lemm:parti}(3).  (c) is correct because of (b).

\begin{defi}[Minimal DD]
Given a set $\Sigma$ of DDs, a DD is minimal if it has a single
attribute DF on the rhs and is not implied by other DDs in $\Sigma$.
\done
\end{defi}

All implication axioms of DDs are given in \cite{diffDep-tods11}. The
ones that are used in this paper will be given shortly and they will
not include the transitivity rule. In other words, our implication is
defined in a restricted way, which is a common shortage in all
level-wise functional dependency (FD) discovery algorithms
\cite{huht99cj-tane-FD-AFD,nov01icdt-discv-EmbFD,yao08dmkd-minFD}.

Minimal DDs are very different from minimal FDs because of attribute
intervals. In FDs, if $A\rar B$, then $AC\rar B$ is not minimal.
However in DDs, if $\aw{A,w_1}\rar \aw{B,w_2}$ holds, the DD
$\aw{A,w_1}\aw{C,w_3}\rar \aw{B,w_{2b}}$ may still be minimal because
$w_{2b}$ can be smaller from $w_2$. When DDs and FDs are compared,
intervals for DDs work like extra attributes for FDs.

The rules below will be used to detect non-minimal DDs. We are
interested in discovering minimal DDs and non-minimal ones will be
ignored.

\begin{lemm}\label{lemm:non-minimal}
If $\aw{X,w_x} \rar \aw{B,w}$ holds, then the following DDs are not
minimal:
\begin{lista}
 \item $\aw{X,w_1}\rar \aw{B,w}$ if $\aw{X,w_x}\succeq \aw{X,w_1}$. (smaller lhs)
 \item $\aw{X,w_x}\aw{Y, w_y} \rar \aw{B,w}$. (extra DF $\aw{Y, w_y}$ making lhs smaller)
 \item $\aw{X,w_x} \rar \aw{B,w_2}$ where $\aw{B,w_2}\succeq\aw{B,w}$. (larger rhs)
 \item If $\aw{B,w_2}\succeq \aw{B,w}$, then DD
$\aw{X,w_x}\aw{B,w_2}\aw{Y,w_y} \rar \aw{C,w_c}$ is implied by
$\aw{X,w_x}\aw{Y,w_y} \rar \aw{C,w_c}$. (lhs reducible)
 \end{lista}
 \end{lemm}

These rules can be easily proved using Lemma \ref{lemm:parti-satisfy}. They lay the
foundation for minimality check in the algorithm proposed later.

\begin{lemm}\label{lemm:combine}
If $\aw{X,w_x}\aw{A,w_1} \rar \aw{B,w}$ and $\aw{X,w_x}\aw{A,w_2}
\rar \aw{B,w}$ hold and $w_1$ and $w_2$ are adjacent, then
$\aw{X,w_x}\aw{A,w_1+ w_2} \rar \aw{B,w}$.
\end{lemm}

The lemma is correct because the two DDs have the same rhs and
because of Lemma \ref{lemm:parti-satisfy}(a). It will be used to
combine DDs so that non-minimal DDs described by Lemma
\ref{lemm:non-minimal}(a) can be removed.

\section{DD satisfaction}\label{sec:satisfy}

In this section, we present results on properties of DD satisfaction
and define approximate satisfaction. The results will guide us to
find satisfied DDs more efficiently.

Consider DD $\aw{X,w_x}\rar \aw{B,w}$ where $B\not\in X$. We would
like to know what $w$ should be if the DD is satisfied. Assume that
two points $p_1$ and $p_2$ represent two tuples in relation $r$ and
they are $w_x$-apart from each other along $X$. Let their
$B$-distance be defined by $abs(p_1[B]-p_2[B])$ where $p[B]$ means
the $B$ coordinate of $p$. The \emph{maximal (minimal resp.)
$B$-distance} for $\aw{X,w_x}$ is the maximal (minimal) $B$-distance
of all the point pairs that are $w_x$-apart along $X$ in the space
$XB$.

\begin{figure}[h]
  \center
  \includegraphics[scale=.9]{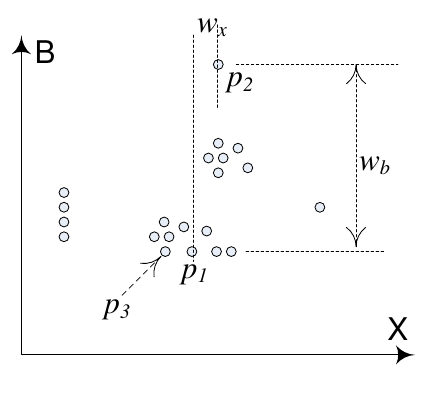}
  \caption{Maximal height $w_b$ of interval $w_x$ \label{f:patch}}
\end{figure}

Figure \ref{f:patch} shows some data points represented by small
circles and a $B$-distance $w_b$ for the interval $w_x$. This
distance is also the maximal $B$-distance for $w_x$. The minimal
$B$-distance is $0$ which occurs between $p_1$ and $p_3$ and some
other pairs. Motivated by the example, we have the following
observation.

\begin{obsv}\label{lemm:max-w}
The DD $\aw{X,w_x}\rightarrow \aw{B,w}$ is satisfied by a relation
$r$ if and only if $left(w)$ is the minimal $B$-distance and
$right(w)$ the maximal $B$-distance for all tuple pairs that are
$w_x$-apart on $X$.
\end{obsv}

The observation can also be formally proved, which is straightforward
and left out.

In many cases, especially when the number of tuples of the relation
is large, $left(w)$ would be 0. The case $left(w)\not =0$ happens
only if there are no two points that are $w_x$-part on a plane
parallel to $X$. One example of such a case is that the distinct
points form vertical lines perpendicular to the plane (like the 4
points on the left in Figure \ref{f:patch}) and the lines are more than
$w_x$-apart. Obviously the $left(w)\not =0$ case is rare.


An implication of this observation is that for any DD
$\aw{X,w_x}\rightarrow \aw{B,w}$, if $w$ is wide enough, the DD is
always satisfied. Let $maxd(B)$ be the maximal distance of attribute
$B$ of all tuple pairs in the relation $r$. We have the following
observation.

\begin{obsv}\label{lemm:max-w}
The DD $\aw{X,w_x}\rightarrow \aw{B,[0,maxd(B)]}$ is always satisfied
without respect to what $\aw{X,w_x}$ is.
\end{obsv}

The observation indicates that for any DF $\aw{X,w_x}$ and an
attribute $B$, we can always find $[0,maxd(B)]$ such that
$\aw{X,w_x}\rightarrow \aw{B,[0,maxd(B)]}$ is satisfied. However such
a DD is not useful because the rhs interval does not indicate any
specific closeness. Thus, in addition to the search for valid
$\aw{X,w_x}$ and pruning the search, which will be presented in the
next section, another problem of the discovery is to find the
tightest $w$ for $\aw{X,w_x}$ and $B$. This is done by following the
partition definitions and Observation 3.1.

We calculate DF partition $F(\aw{X,w_x})$ and the attribute partition
${\cal F}(B)$ which contains ordered partitions of DFs of the single
attribute $B$. We then search ${\cal F}(B)$ from right to left for
the first DF partition $F(\aw{B,w_j})$ that overlaps with
$F(\aw{X,w_x})$. The $right(w_j)$ is what $right(w)$ should be. In
the same way, we search ${\cal F}(B)$ from left to right for the
first DF partition $F(\aw{B,w_i})$ that overlaps with
$F(\aw{X,w_x})$. The $left(w_i)$ is what $left(w)$ should be.
Formally,
\begin{align}
    \begin{aligned}\label{eq:w}
  left(w)= min\{left(w_i) \ |\  & F(\aw{B,w_i})\in \calF(B) \land \\
  &    F(\aw{B,w_i})\cap   F(\aw{X,w_x})\not=\phi\} \\
  right(w)= max\{right(w_j) \ |\ & F(\aw{B,w_j})\in \calF(B) \land \\
  &    F(\aw{B,w_j})\cap F(\aw{X,w_x}) \not=\phi\}
       \end{aligned}
\end{align}

Because the DF partitions of $\calF(B)$ are ordered by $w_i$,
calculating the set intersections from the two ends of $\calF(B)$ is
efficient.
 \ \\

We now show the motivation for approximate satisfaction of DDs. Given
two DDs $f_1=\aw{X,w_x}\rightarrow \aw{B,[0,d_1]}$ and
$f_2=\aw{X,w_x}\rightarrow \aw{B,[0,d_2]}$ with the same lhs, if
$d_2$ is much smaller than $d_1$, then, $f_2$ is much more
interesting than $f_1$ as the tuples satisfying the lhs are closer on
$B$. Closeness is the soul of DDs. Unfortunately, isolated points
like $p_2$ in Figure \ref{f:patch}, have significant negative impact
on the usefulness of DDs. As shown in the figure, the $w_b$ must
cover the distance of the pair $p_1$ and $p_2$ for the DD to be
satisfied. In contrast, if $p_2$ did not exist, the $w_b$ would be
much smaller and covers just the dense patch of pairs. For this
reason, we define approximate DDs.

\begin{defi}[Approximate differential dependency satisfaction]
Given a large number $\epsilon$ in [0,1] called the approximate
satisfaction threshold, the DD $\aw{X,w_x}\rar \aw{B, w}$ is
approximately satisfied ($\epsilon$-satisfied) if, for all tuple
pairs satisfying $\aw{X,w_x}$, at least $\epsilon$ fraction of them
with lowest distance satisfy $\aw{B,w}$. \done
\end{defi}

The definition can be easily adapted if the majority of tuple pairs
have large distances and a few outliers have very small distances.
The $\epsilon$-satisfaction is different from the confidence of
associations because the order is involved in the definition.

\begin{defi}[Interestingness]
Given DD $\sigma=\aw{X,w_x}\rar \aw{B,w}$ and the relation $r$, the
\emph{support} to $\sigma$, denoted by $supp(\sigma)$, is the ratio
of the number of all tuple pairs satisfying $\aw{X,w_x}\aw{B,w}$ over
the number of all tuple pairs of $r$. Let $maxd(B)$ be the maximal
distance of attribute $B$ in $r$. The \emph{interestingness} of
$\sigma$ is defined to be
\[intr(\sigma)\ =\ {\ \ \ supp(\sigma)\ \ \  \over{w \over maxd(B)}}\]
 \done
\end{defi}

This definition gives each discovered DD a number. The larger the
number is, the more interesting the DD is. A DD discovered with
higher support has higher interestingness. A DD discovered having a
narrower rhs interval also has higher interestingness. Importantly,
the definition implies that the DDs with fewer attributes on the lhs
are more interesting because it is often the case that the DDs with
less attributes on the lhs are satisfied by more tuple pairs and
therefore have more support than DDs with more attributes on the lhs.

Interestingness can have a tie in two cases: the case where the
support is small and the rhs interval is narrow and the case where
the support is large and the rhs interval is wide. Both cases are not
interesting to the discovery and such a tie does not matter.

\subsection{Outliers}
Finally in this section, we show the connection between outliers,
fully satisfied DDs and approximately satisfied DDs. Let
$f_1=\aw{X,w}\rar \aw{B,w_1}$ be a fully satisfied DD from a relation
$r$ and $f_2=\aw{X,w}\rar \aw{B,w_2}$ be an $\epsilon$-satisfied DD.
Obviously $w_2\le w_1$. The ratio \[rr={w_1-w_2 \over w_1}\]
indicates the amount of interval width reduced by $\epsilon$.
\emph{If $rr$ is smaller than $\epsilon$, the chance for the
existence of outliers for $\aw{X,w}\rar \aw{B,w_1}$ is very high.}
This becomes an effective mechanism for outlier detection. By
applying this mechanism to data cleaning, possible outlier tuples can
be identified, which makes it possible for the verification and cleaning
work to be applied to the right point.

\section{DD discovery}\label{sec:disc}

In this section, we present our DD discovery algorithm. With DDs, for
any $\aw{X,w_x}$ and $B$, an interval $w$ can always be found so that
$\aw{X,w_x} \rar \aw{B,w}$ is satisfied. This property of DDs
determines that the discovery of DDs is very different from the
discovery of functional dependencies.

Our way of discovering DDs contains the processes of generating lhs, calculating the rhs
interval $w$, checking minimality and pruning.

To generate lhs of DDs, we assume that each attribute $B$, with the
maximal distance value $maxd(B)$, has a \emph{user-selected
interesting distance range} $[0,ur(B)]$ ($ur(B)\le maxd(B)$), and a
sequence of distance intervals $w_1,\ddd,w_{k-1},w_k$ where $w_{i}$
($i=1,\ddd,k-1$) and $w_{i+1}$ are adjacent, $right(w_{k-1})=ur(B)$,
and $w_k=[ur(B),maxd(B)]$. That is, $w_1,\ddd,w_{k-1}$ cover the
whole user-selected interesting range and $w_k$ includes all the
remaining distances. For example, if the maximum distance interval of
attribute $Age$ is 55, and the interesting distance range of $Age$ is
$[0,40]$, then the sequence of intervals of $Age$ can be $[0,5],
[6,10], \ddd, [36,40],[41,55]$. These intervals are called base
intervals.

The sizes of the base intervals can be small or large. Smaller
intervals will lead to more accurate DDs to be discovered. However,
they also mean a larger $k$, the number of intervals, which has
exponential impact on the complexity of computation. Fortunately in
most applications, only small distances are interesting. For example,
an $Age$ distance over 40 means that a young employee and a retiring
employee are compared. Such a comparison would not be very useful in
many cases. In the same way, in a data quality application in which
typos are to be detected, an editing distance over 10 between two
words may not be helpful. This is because such a 'large' distance should not be
considered for typos.  In the case of applications where large distances
are more interesting, large interval sizes can be used to
cover more small distances while small interval sizes can be used to
cover large distances. We leave the user-selected interesting range
and the interval sizes to be decided by domain experts.

\subsection{Generating lhs}\label{sec:gen_lhs}
The process of generating lhs produces nodes for a lattice. A node is
a tuple $(v, F(v), dds(v))$ where $v=\aw{X,w}$ is a differential
function, $F(v)=F(\aw{X,w})$ is the tuple pair partition for $v$, and
$dds(v)=\{\aw{Y_1,w_{y1}}\aw{B_1,w_1}, \ddd,
\aw{Y_k,w_{yk}}\aw{B_k,w_k}\}$ is the set of differential functions
for both sides of the satisfied DDs $\aw{Y_i,w_{yi}}\rar
\aw{B_i,w_i}$ such that $w_i$ is a base interval of $B_i$ and
$\aw{Y_i,w_{yi}}\subseteq v$. The $\subseteq$ symbol indicates that
$dds(v)$ will be carried from level to level. $F(v)$ is used to test
partition containment for satisfaction, $dds(v)$ is used to detect
reducible DDs.

Level-1 nodes are constructed by single attribute differential
functions. Each attribute $B$ and its base intervals $w_1,\ddd, w_k$
form $k$ differential functions $\aw{B,w_1},\ddd,\aw{B,w_k}$. The
differential functions of all attributes form the first level nodes.
Let $v$ be such a node. We set $dds(v)=\phi$ and calculate $F(v)$ as
defined in Equation (\ref{eq:F}).

\begin{figure}[h] \label{f:lattice}
  \center
  \includegraphics[scale=.8]{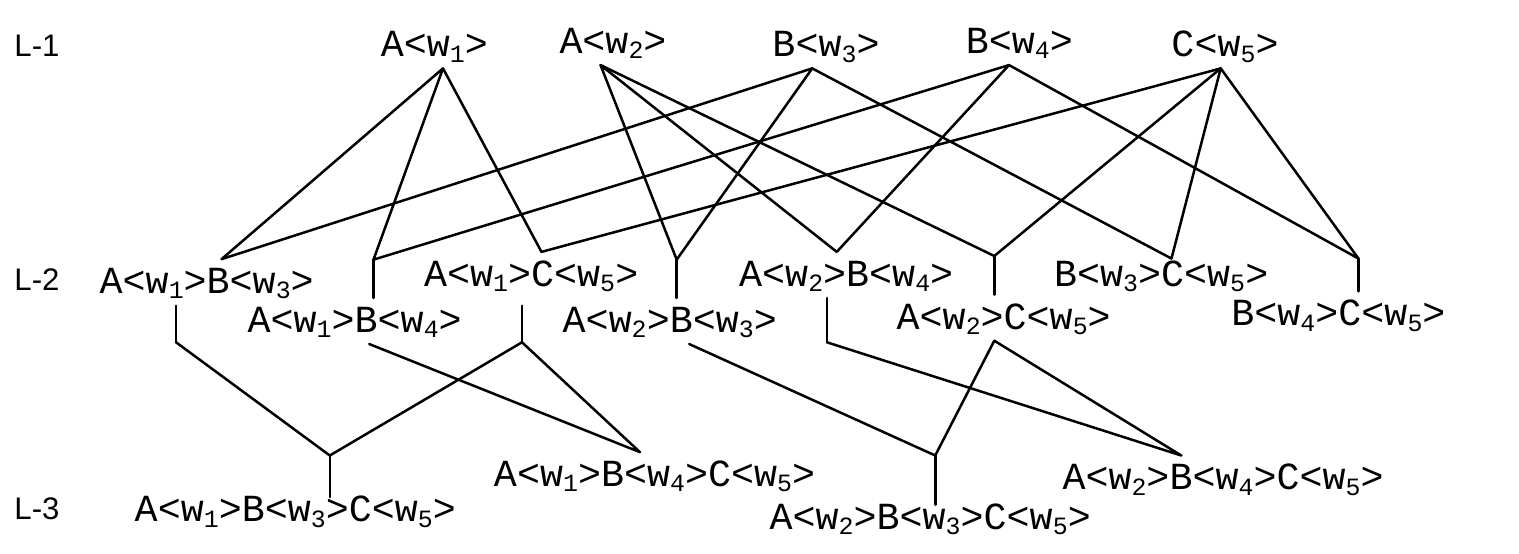}
  \caption{An example lattice}
\end{figure}

A Level-$i$ ($i\ge 2$) node is joined from two Level-($i-1$) nodes if
they share $(i-2)$ preceding single attribute differential functions
and their tailing single attribute differential functions are of
different attributes. For example, the node
$\aw{A,w_1}\aw{B,w_2}\aw{C,w_3}\aw{D,w_4}$ at Level-4 is joined from
$\aw{A,w_1}\aw{B,w_2}\aw{C,w_3}$ and $\aw{A,w_1}\aw{B,w_2}\aw{D,w_4}$
at Level-3. For opposite examples, the join of any pair of the three
Level-3 nodes $\aw{A,w_1}\aw{B,w_2}\aw{C,w_2}$,
$\aw{A,w_1}\aw{B,w_2}\aw{C,w_3}$, and
$\aw{A,w_1}\aw{C,w_2}\aw{D,w_3}$ will not produce a Level-4 node. A
special case of this principle is that a Level-2 node is combined
from two Level-1 nodes if they have different attributes. Figure
\ref{f:lattice} shows the lattice with three attributes $A$, $B$, and
$C$. Each of the fist two has two intervals and $C$ has one interval.

Assume that the relation has $m$ attributes and attribute $A_i$ has
$k_i$ intervals. Let $x=k_1+\ddd+k_m$. The number of nodes in the
lattice is $C(x,1)+C(x,2)+\ddd+ C(x,m)$ where $C(x,k)$ is the
combination of $k$ elements out of $x$ total elements.

We note that the lattice generated here is different from the Apriori
lattice \cite{agra94vldb-min-asso} used in FD discovery
\cite{liu10tkde-FDd}. In the lattice for FD discovery, each edge
represents a candidate FD, all nodes at a level has the same number
of in-coming edges and the same number of out-going edges, and the
last level has only one node. In our lattice for DD discovery, edges
represent how nodes are generated, and they do NOT represent
candidate DDs. The nodes at the same level have two in-coming edges
showing the derivation but may have different number of out-going
edges. The last level has multiple nodes in general. The candidate
DDs are then derived from each of the nodes.

\subsection{Calculating rhs interval $w$}\label{sec:cal_rhs}
The components of a node are derived from the joining nodes of the
previous level. Let node $v=\aw{X,w_x}$ be joined from the two nodes
$v_1=\aw{Y,w_y}$ and $v_2=\aw{Z,w_z}$ at the previous level. Then
$F(v)=F(v_1)\cap F(v_2)$, and $dds(v)=dds(v_1)\cup dds(v_2)$.

The candidate DDs of a node $v=\aw{X,w_x}$ are generated using $v$ as
the lhs and the attributes that are not in $v$ as the rhs. The
intervals of the rhs attributes are represented by '$w$' which are to
be decided. Let $v=\aw{A,w_a}\aw{B,w_b}$ and $R=\{A,B,C,D\}$. Then
the candidate DDs are $v\rar \aw{C,w_c}$, $v\rar \aw{D,w_d}$.

Deciding the rhs interval $w$ of a candidate DD follows Equation
(\ref{eq:w}). If $right(w)>ur(B)$ where $ur(B)$ is upper limit of the
user-selected interesting range of $B$, the DD is trivial. Otherwise,
it is not. A non-trivial DD $f=\aw{X, w_x} \rar \aw{B,w}$ is added to
$\Sigma$, the set of all discovered DDs, if it is not implied by
other DDs that are already in $\Sigma$. If $w$ equals to a specific
base interval $w_i$, then $\aw{X,w_x}$ is added to $dds(v)$, which
will be used to detect reducible non-minimal DDs.

\blankline When approximate satisfaction is used with threshold
$\epsilon$, Equation (\ref{eq:w}) is modified to the following to
decide the interval $w$. Note that both formulae takes minimal.
\begin{align}
\begin{aligned}
  left(w)&=min\{left(w_i) \ | \ F(\aw{B,w_i})\in \calF(B) \land c_1=True \}\\
  right(w)&=min\{right(w_i) \ | \ F(\aw{B,w_i})\in \calF(B) \land c_1=true  \land c_2=true\}
\end{aligned}
\end{align}
 where \[ c_1=  F(\aw{B,w_i})\cap F(\aw{X,w_x}) \not=\phi\]
  \[c_2={ \Sigma_{j=i}^k|F(\aw{B,w_j})\cap
F(\aw{X,w_x})| \over |F(\aw{X,w_x})|}< (1-\epsilon)\]

\noindent These equations are implemented in the following function where $F(\aw{B,w_i})$
is the $i$-th element of $\calF(B)$.

\noindent\rule{\textwidth}{0.2pt}\vspace{-2em}
\begin{func}\label{func:w}
\begin{algorithmic}[1]\fontfamily{pcr}\selectfont
\renewcommand{\algorithmicrequire}{\textbf{FindRhs}\hspace{-1.5ex}}
  \REQUIRE ($F(\aw{X,w_x})$, $\calF(B)$, $\epsilon$) as interval:
  \STATE $cnt=0$;
 \FOR{$i=k,\ddd,1$}
    \STATE  $F=F(\aw{X,w_x})\cap F(\aw{B,w_i})$,   $cnt += |F|$ \\
          if ${cnt\over |F(\aw{X,w_x})|}> (1-\epsilon)$: $r=right(w_i)$, break
  \ENDFOR
 \FOR{$i=1,\ddd,k$:}
    \STATE  $F=F(\aw{X,w_x})\cap F(\aw{B,w_i})$,   $cnt = |F|$ \\
          if $cnt>0$: $l=left(w_i)$, break
  \ENDFOR
  \STATE return $[l,r]$
\end{algorithmic}
\vspace{-.8em}\noindent\rule{\textwidth}{0.2pt}
\end{func}

\subsection{Pruning}\label{sec:pruning}
The lattice generation and $w$ calculation guarantee that all the
single attribute differential functions on the lhs are with only base
intervals and all rhs intervals are minimal. We discuss the rules
of Lemma \ref{lemm:non-minimal}. The condition of rule (a) will never
be met because $w_x$ is a base interval and the smaller interval
$w_1$ does not exist. The condition $\aw{B,w_2}\succeq \aw{B,w}$ of
rule (c) will never be met because for the same lhs in
$\aw{X,w_x}\rar \aw{B,w}$ and $\aw{X,w_x}\rar \aw{B,w_2}$, we always
find $ \aw{B,w}$, not $\aw{B,w_2}$. The condition $\aw{B,w2}\succeq
\aw{B,w}$ of rule (d) needs to be specialized to
$\aw{B,w_2}=\aw{B,w}$ because the intervals on the lhs can only be
base intervals. Thus in minimality check, we only need to consider
rules (b) and (d) with the condition of $(d)$ changed to
$\aw{B,w_2}=\aw{B,w}$.

Following the lattice generation and the discussion above, we have
the following lemma.

\begin{lemm}\label{lemm:reduce}
If $\aw{X,w_x}\rar \aw{B,w}$ where $w$ is a base interval, all nodes
containing $\aw{X,w_x}\aw{B,w}$ can be pruned.
\end{lemm}

\noindent\bfit{Proof:} Consider DD $f_1=\aw{X,w_x}\aw{B,w}\aw{Y,w_y}
\rar \aw{C,w_c}$. By Lemma \ref{lemm:non-minimal}(d), $f_1$ can be
reduced to $f_2=\aw{X,w_x}\aw{Y,w_y} \rar \aw{C,w_c}$. $f_2$ involves
less attributes and it must have been checked at a previous level.
Consequently nodes like $\aw{X,w_x}\aw{B,w}\aw{Y,w_y}$ can be pruned
from the lattice. \done

We note that we can use $\aw{X,w_x}\rar \aw{B,w}$ to prune nodes
containing $\aw{X,w_x}\aw{B,w}$, but we cannot avoid computing $w_2$
for $\aw{X,w_x}\aw{A,w_a}\rar \aw{B,w_2}$. The reason is that $w_2$
may be narrower than $w$ and if this is the case,
$\aw{X,w_x}\aw{A,w_a}\rar \aw{B,w_2}$ is a true minimal DD. This is
different from the case in FD discovery \cite{huht99cj-tane-FD-AFD}.
In FD discovery, if $X\rar B$, the check of $XA\rar B$ is not
necessary. Furthermore, if every attribute $B$ in the node $X$ is
determined non-trivially by a subset of $X$, the node can be pruned.
The two FD discovery pruning rules do not apply to DD discovery.

\blankline We also use the support to a node $\aw{X,w_x}$ to prune
the nodes from the lattice. The support is the ratio of tuple pairs
satisfying $\aw{X,w_x}$ over all tuple pairs and is calculated by
$supp(\aw{X,w_x})={|F(\aw{X,w_x})| \over |r|}$. If the support is
lower than a threshold $\delta$, no DD will be calculated for the
node. The reason for using a support threshold is that we do not want
DDs satisfied by only a few tuple pairs to be discovered. Such DDs do
not reflect typical relationships.

The following lemma is used in many association rule mining works.

\begin{lemm}\label{lemm:supp}
Given a support threshold $\delta$ and a node $\aw{X,w}$, if
$supp(\aw{X,w})<\delta$, $supp(\aw{X,w}\aw{Y,w_y})<\delta$ for any
$\aw{Y,w_y}$. All nodes containing $\aw{X,w}$ can be pruned.
\end{lemm}
The lemma is correct because $F(\aw{X,w}\aw{Y,w_y})= F(\aw{X,w})\cap
F(\aw{Y,w_y})$, $|F(\aw{X,w}\aw{Y,w_y})| < |F(\aw{X,w})|$, so
$supp(\aw{X,w}\aw{Y,w_y}) < supp(\aw{X,w})$.

\begin{algorithm}
\caption{MinDD} \label{alg:mindd}
 \algsetup{ indent=3ex, linenosize=\small,linenodelimiter=. }
\begin{algorithmic}[1]   
\renewcommand{\algorithmicrequire}{\textbf{Inp:}}
\renewcommand{\algorithmicensure}{\textbf{Outp:}}
 \REQUIRE relation $r$ on schema $R$, ${\cal W}_A=[w_1,\ddd,w_k]$ for every attribute $A\in R$,
 satisfaction threshold $\epsilon$ and support threshold $\delta$
 \ENSURE the set of minimal DDs $\Sigma$
 \STATE   Calculate $F(\aw{A,w})$ for all $A\in R$ and all $w\in {\cal W}_A$; \\
  Calculate $\calF(A)$ for all $A\in R$.
 \STATE   Create level-1 nodes: for each $A\in R$ and for each interval $w\in {\cal
  W}_A$,
   \{let $v=\aw{A,w}$;
     \ \  if $|F(\aw{A,w})|>0$: add node  $(v, F(v)=F(\aw{A,w}), dds(v)=\phi)$ to $L_1$ \}
 \FOR{level $i=1,\ddd,|R|$}
    \FOR{node $v\in L_i$}
       \STATE if $\ex \aw{Y,w_y}\in dds(v)(\aw{Y,w_y}\in v)$: $v$ is reducible, remove $v$ from $L_i$, continue\\
       \FOR{attribute $B\in R$ and $B\not\in attr(v)$}
         \STATE let candidate DD be $f=v\rar \aw{B,w}$ \\
         \STATE // calculate rhs: \\
         \STATE   $w=FindRhs(F(v), \calF(B),\epsilon)$ [Function \ref{func:w}] \\
         \STATE if $w \not\subseteq [0,ur(B)]$: continue; \\
         \STATE if $ChkImply(f,\Sigma)=false$: add $f$ to $\Sigma$ \\
         \STATE if $w$ is a base interval, add $v\aw{B,w}$ to $dds(v)$.
        \ENDFOR \ attribute $B$
     \ENDFOR \ node $v$
     \STATE // build nodes for next level \\
     \STATE let $v_1=\aw{X,w_x}\aw{A,w_a}$ and $v_2=\aw{X,w_x}\aw{C,w_c}$ ($A\not=C$) be nodes in $L_i$
     \STATE $v=\aw{X,w_x}\aw{A,w_a}\aw{C,w_c}$,  $F(v)=F(v_1)\cap F(v_2)$,
      $dds(v)=dds(v_1)\cup dds(v_2)$ \\
      \STATE if $|F(v)|\ge \delta$ (support pruning): add $(v,F(v), dds(v))$ to $L_{i+1}$\\
  \ENDFOR \ level $i$
  \STATE return $\Sigma$
\end{algorithmic}
\end{algorithm}

\subsection{The Algorithm}\label{subsec:algorithm}

Our algorithm is called $MinDD$ and is given in Algorithm
\ref{alg:mindd}. It builds the lattice in the breadth-first manner
and for each node to be added to the lattice, it calculate DDs for
the nodes. Line 1 of the algorithm calculates tuple pair partitions
for differential functions and for single attributes. Line 2 builds
the first level nodes. Line 5 uses $dds(v)$ to check reducibility and
to prune nodes based on Lemma \ref{lemm:reduce} from the current
level. Lines 7-12 derive candidate DDs and determines their rhs
intervals. In Line 10, the algorithm uses the user-selected interesting
range of the rhs attribute $B$ to filter the found DD: if it is not
in the interesting range, it is ignored. Line 11 uses the function
$ChkImply()$ to check to see if the DD is implied by other DDs in
$\Sigma$. This function will be described in detail later. Line 12
maintains the differential functions of satisfied DDs of the node.
Lines 15-17 builds nodes for the next level and Line 18 prunes the
node if it does not have enough support.

The complexity of this algorithm is analyzed as follows. The lattice
has $2^{\Sigma w}$ nodes where $\Sigma w$ is the sum of the number of
intervals of all attributes. The partition of a node is the product
of the partitions of the two participating parent nodes. This takes
${({|r|^2\over 2})^2}$ operations where ${{|r|^2\over 2}}$ is the
number of tuples of the distance relation of $r$. Thus, the complexity of
the algorithm is $O({({|r|^2\over 2})^2}2^{\Sigma w})$.

\subsection{Detection of implication of DDs}
We design a tree structure, called a DD-tree, to store all DDs having the same rhs
differential function, and use the tree to check whether a DD is implied. The root node
of the tree is the rhs $\aw{B,w_B}$ of the DDs. Other nodes are single attribute
differential functions in the lhs of these DDs. Child nodes of a node are sorted by their
attributes and their intervals. A path
$\aw{B,w_{B}}/\aw{A_1,w_{1}}/\aw{A_2,w_{2}}/\ddd/\aw{A_j,w_{j}}$ represents the DD
$\aw{A_1,w_{1}}\aw{A_2,w_{2}}\ddd\aw{A_j,w_{j}}\rar \aw{B,w_{B}}$.

\begin{defi}[path prefix] A path $\aw{B,w_{B}}/\aw{A_1,w_{1}}/\aw{A_2,w_{2}}/\ddd/\aw{A_j,w_{j}}$
is a prefix of path
$\aw{B,w_{B}}/\aw{\bar{A}_1,\bar{w}_{1}}/\aw{\bar{A}_2,\bar{w}_{2}}/\ddd/\aw{\bar{A}_k,\bar{w}_{k}}$
if $j<k$ and for each $i=1\ddd j (\aw{A_i,w_{i}}\succeq\aw{\bar{A}_i,\bar{w}_{i}} )$.
\end{defi}

\begin{lemm}
A DD $f_2$ is implied by DD $f_1$ if the path of $f_1$ is a prefix of
the path of $f_2$.
\end{lemm}

This lemma is correct because of Lemma \ref{lemm:non-minimal}.

\begin{defi}[non-redundant] A DD tree is non-redundant if no node has two or more child nodes of
the same differential functions.
\end{defi}

Assume that a hash table $h(\aw{B,w_B}, tr)$ is created to store the
rhs differential functions like $\aw{B,w_B}$ and their trees. With
the hash table, the following function checks whether the given DD is
implied by previously found DDs in the way specified by Lemma
\ref{lemm:non-minimal} where the $Combine()$ function will be
introduced later.

\noindent\rule{\textwidth}{0.2pt}\vspace{-2em}
\begin{func}\label{func:imply}
\begin{algorithmic}[1]\fontfamily{pcr}\selectfont
\renewcommand{\algorithmicrequire}{\textbf{ChkImply}\hspace{-1.5ex}}
  \REQUIRE ( DD $f$ ) as bool:
  \STATE assume $f=\aw{A_1, w_{A_1}}\ddd \aw{A_j, w_{A_j}} \rar \aw{B,w_B}$
  \STATE let $h=h(\aw{B,w_B}, tr)$ be the hash table.
  \STATE retrieve $tr$ using $\aw{B,w_B}$ from $h$.
  \STATE  convert $f$ to path $p=\aw{B,w_B}/\aw{A_1, w_{A_1}}/\ddd /\aw{A_j, w_{A_j}}$ \\
      \IF{ $tr$ contains path $q$ s.t. $q$ is a prefix of $p$:}
            \STATE return false
       \ELSE \STATE add $\aw{A_1, w_{A_1}}/\ddd /\aw{A_j, w_{A_j}}$ to the root of
       $tr$,\\
          $tr$ is now redundant, $Combine(tr,p)$.
      \STATE return true
       \ENDIF
\end{algorithmic}
\vspace{-.8em}\noindent\rule{\textwidth}{0.2pt}
\end{func}

The $Combine()$ function which is defined in Procedure
\ref{proc:combine} implements Lemma \ref{lemm:combine}. Two child
nodes $v_1=\aw{A,w_1}$ and $v_2=\aw{A,w_2}$ of a node is combinable
if $v_1$ and $v_2$ have identical child trees. The child trees of
node $v$ is denoted by $children(v)$. The combination extends the
interval of $v_1$ to $w_1+ w_2$ and deletes $v_2$.

\noindent\rule{\textwidth}{0.2pt}\vspace{-2em}
\begin{proc}\label{proc:combine}\
\begin{algorithmic}[1]\fontfamily{pcr}\selectfont
\renewcommand{\algorithmicrequire}{\textbf{Combine}\hspace{-1.5ex}}
  \REQUIRE (dd-tree $tr$, path $p=\aw{B,w_B}/\aw{A_1, w_{A_1}}/\ddd /\aw{A_j, w_{A_j}}$):
  \STATE locate the last node $v_1=\aw{A_j, w_{A_j}}$ of $p$ in $tr$.
 \FOR{$v_1=\aw{A_1, w_{A_1}},\ddd ,\aw{A_j, w_{A_j}}$}
     \STATE let $x$ be parent of $v_1$
    \STATE find another child node $v_2=\aw{A, w_2}$ of $x$   \\
    \STATE  {\bf if} $v_2==null$ or $children(v_1)!=children(v_2)$: break
          \STATE replace $\aw{A, w_{A}}$ by $\aw{A, w_{1}+ w_{2}}$;
                  delete $v_2$.
  \ENDFOR
\end{algorithmic}
\vspace{-.8em}\noindent\rule{\textwidth}{0.2pt}
\end{proc}

\noindent As an example, given DDs
\begin{tab}
 $f_1=\aw{A, w_1}\aw{C,w_2}\aw{D, w_3}\rar \aw{B, w}$, \\
 $f_2=\aw{A, w_1}\aw{C, w_3}\aw{D, w_3}\rar \aw{B, w}$, and \\
 $f_3=\aw{A, w_1}\aw{C, w_3}\aw{D, w_4}\rar
\aw{B, w}$ checked by $ChkImply()$ in order,
 \end{tab}
  the tree $tr$ with root $\aw{B, w}$ have the
following two child paths $\aw{A, w_1}/\aw{C, w_2+w3}/\aw{D, w_3}$
and $\aw{A, w_1}/\aw{C, w3}/\aw{D, w_4}$.

\blankline We use the following lemma to show that DDs with
implication relationship will not be put into different DD-trees.

\begin{lemm}\label{lemm:implyComplete}
If DDs $f_1$ and $f_2$ are discovered by Algorithm \ref{alg:mindd} and $f_2$ is implied
by $f_1$, both DDs will be directed to the same DD-tree.
\end{lemm}

\noindent{\bfit Proof:} If $f_1=\aw{X,w_x}\rar \aw{B,w}$, based on
previous discussion (Lemma \ref{lemm:non-minimal} and Section
\ref{sec:pruning}), DD implication happens in the following cases.
 (1) $\aw{X,w_x}\aw{B,w}\aw{A,w_a}\rar
\aw{C,w_c}$ is reducible. (2) $f_2=\aw{X,w_x}\aw{A,w_a}\rar \aw{B,w}$
is implied.

Case (1) does not happen because the algorithm uses $dds(v)$ to
filter such DDs. Case (2) is the only way DD implication happens.
Because $f_2$ and $f_1$ have the same rhs, they are put into the same
DD-tree.  \done

The lemma implies that the implication detection is complete and no
implied DDs will be output by the algorithm.

\section{Sampling and errors}\label{sec:sampling}

The discovery algorithm has the complexity factored by $|r|^4$ (see
subsection \ref{subsec:algorithm}) for relation $r$. This indicates that the
discovery from large data sets is not possible. For this reason, we
use sampling when data sets get large.

The sample size has to be determined before sampling can be
conducted. We note that sampling for dependency discovery is
different from sampling for many statistical studies and the
difference is the number of results that can be observed from a
sample. For example, in the study of the percentage of heads to
appear in coin toss, each toss gets a result (true or false). Thus
if in a sample a coin is tossed for 380 times, 380 results will be
observed. For such problems in statistics, well-developed formulas
for determining the sample size are available.

In dependency discovery, however, the satisfaction must be
calculated based on the whole data set, not on individual tuples. In
other words, if a sample contains 380 tuples drawn from the original
data set, only one result (true or false) about the satisfaction of
a DD can be observed. Because of this difference, sample sizes in
dependency discovery must be carefully studied.

Consider a data set $r$, its distance relation $\bar{r}$, and a
candidate DD $f$. Assume that $\bar{r}$ has $N$ total tuples and
$M_f$ of them support $f$. The sample must be taken without
replacement because each tuple in $\bar{r}$ is used only once in
testing the satisfaction of $f$. When a sample $\bar{r}_s$ of $N_s$
tuples are drawn randomly from $\bar{r}$ with no replacement, the
probability for $\bar{r}_s$ to have exact $k_f$ supporting tuples
follows the hypergeometric distribution
\[P(N,M_f, N_s,k_f)={\combin{M_f}{k_f} \combin{N-M_f}{{N_s-k_f}}\div \combin{N}{{N_s}}}\]
where the notation $\combin{x}{{y}}$ is the number of
$y$-combinations from $x$ elements.

The sampling process causes change to: (1) the support to the DD $f$
and (2) the rhs interval of $f$. We firstly analyze the
change to the support.

Given a support threshold $\theta$, two types of erroneous DDs,
namely {\it missed-wanted} DDs and {\it found-unwanted} DDs, may
happen because of sampling. A {\bf wanted} DD is one with a support
equal to or higher than the threshold $\theta$ in $\bar{r}$ (before
sampling) and an {\bf unwanted} DD is one with a support lower than
the threshold $\theta$ in $\bar{r}$. A DD $f$ is {\bf missed-wanted}
({\bf missed} for short) if the number of its supporting tuples in
$\bar{r}$ is more than the threshold ($M_f\ge \theta N$), but the
number of its supporting tuples in the sample $\bar{r}_s$ is less
than the threshold ($<\theta N_s$). An example of such a DD is $f_2$
in Table \ref{tbl-err-missed wanted}. The probability for a wanted
$f$ to be missed is
\begin{equation}\label{eq:missed wanted} P(N,M_f,N_{s},k_f < \theta N_{s})=
\sum_{k_f=0}^{\theta N_{s}-1}
 P(N,M_f,N_{s},k_f)
 \end{equation}
We like this probability to be small.

\begin{table}[h]
\center \caption{Missed DDs and unwanted DDs. (b) is a sample from
(a). The set of DDs found from (a) is listed in the middle and the
set of DDs from (b) are listed below (b).} \label {tbl-err-missed
wanted}
\begin{tabular}{|l|l|l}
 \multicolumn{2}{c}{(a): $\bar{r}$} \\ \hline
 {\bf A} & {\bf B}     \\ \hline
  0 & 1    \\ \hline
  0 & 1   \\ \hline
  0 & 1   \\ \hline
  1 & 2   \\ \hline
  2 & 3    \\ \hline
  2 & 3   \\ \hline
\end{tabular}
\begin{tabular}{l}
 DDs on $\bar{r}$: ($\theta=1/3$)\\
 $f_1=\aw{A,0}\rar\aw{B,1}$\\
 $f_2=\aw{A,2}\rar\aw{B,{3}}$ \\
\end{tabular}
\begin{tabular}{|l|l|}
 \multicolumn{2}{c}{(b): $\bar{r}_s$ (sample rate=1/3)}  \\ \hline
 {\bf A} & {\bf B}     \\ \hline
  0 & 1   \\ \hline
  1 & 2   \\ \hline
  \multicolumn{2}{l}{DDs on $\bar{r}_s$: (same $\theta$)}\\
  \multicolumn{2}{l}{$f_1=\aw{A,0}\rar\aw{B,1}$}\\
  \multicolumn{2}{l}{$f_3=\aw{A,1}\rar\aw{B,2}$} \\
  \multicolumn{2}{l}{$f_2$ missed. $f_3$ is unwanted .}
\end{tabular}
\end{table}

From this equation, we can derive the sample size $N_{s}$ if we set
the probability to a specific value. An explicit formula for getting
the sample size $N_{s}$ seems not easy to get. However, it is easy to
program a calculator to determine $N_{s}$ given all other parts of
the formula.

Assume that $\bar{r}$ has $N=10000$ tuples, $M_f=10$ of them support
$f$, and $\theta=0.0005$. If we set the chance for $f$ to be missed
to $P(N,M_f,N_{s},k_f < \theta N_{s})=0.05$, the sample size $N_{s}$
needs to be $3941$. If we choose the smallest threshold:
$\theta=0.0001$ for the specific $N$, for the same $M_f$ and same
$P$, the sample size must be $2588$. The smaller threshold leads to
smaller sample size.

If we reduce the number of supporting tuples $M_f=1$ and keep
$\theta=0.0001$ and $P(N,M_f,N_{s},k_f < \theta N_{s})=0.05$, the
sample size must be $9501$. Obviously the proportion of supporting
tuples and the threshold control the sample size. The smaller the
number of supporting tuples or the larger the threshold, the larger
the sample size has to be.

A {\bf found-unwanted } ({\bf unwanted} for short) DD $f$ is opposite
to a missed wanted DD. $f$ is found-unwanted  if it has ${M_{f} <
\theta N}$ number of supporting tuples in $\bar{r}$ but has $\theta
N_s$ or more supporting tuples in the sample $\bar{r}_s$. An example
of a found-unwanted  DD is $f_3$ in Table \ref{tbl-err-missed
wanted}. The probability for this to happen is

\begin{equation}\label{eq:found-unwanted }
P(N,M_f,N_{s},k_f \ge \theta N_{s})= 1-
\sum_{k_f=0}^{\theta N_{s}-1}P(N,M_f,N_{s},k_f)
\end{equation}
We like this probability to be small. By comparing Formulas
(\ref{eq:missed wanted}) and (\ref{eq:found-unwanted }), we find that
the two are complementary. This implies that irrespective of the sample
size (small or large), the chance of one of the errors is high as
$M_f$ in both cases is close and small. Thus, the analysis of support
change caused by sampling does not help with the determination of the
sample size.

The above of analysis is on the possibility for one DD to become an
error. The total number of erroneous DDs discovered from a sample
relates the total number of DDs from the distance relation, their
supports, and their probabilities. Let $\Sigma_a$ (resp. $\Sigma_b$)
be the set of all wanted DDs (resp. unwanted DDs) on the distance
relation $\bar{r}$, $f_i$ a DD in $\Sigma_a$ or $\Sigma_b$ with
$M_{f_i}$ supporting tuples in $\bar{r}$ and $k_{f_i}$ supporting
tuples in the sample, and $\theta$ the support threshold. The number
$E_m$ of DDs missed and the number $E_{uw}$  of unwanted DDs found
from the sample are given in Formula (\ref{eq:ttl-err-dds }) below.

\begin{align}
    \begin{aligned}
    E_{m}= \sum_{f_i\in \Sigma_a} P(N,M_{f_i},N_{s},k_{f_i} \ge \theta
    N_{s}),\ \ \  M_{f_i}< \theta N \\ \label{eq:ttl-err-dds }
    E_{uw}=
    \sum_{f_i\in \Sigma_{b}} P(N,M_{f_i},N_{s},k_{f_i} < \theta
    N_{s}),\ \ \ M_{f_i} \ge \theta N \\
    \end{aligned}
\end{align}
Note that the $\Sigma_b$ contains DDs with low support and such DDs
are much more in number than the wanted DDs in $\Sigma_a$.
Consequently $E_{uw}>>E_{m}$. This has been confirmed by our
experiments.

A further type of error caused by sampling is the reduction of rhs
intervals. This can be demonstrated by the data in Table
\ref{tbl-err-reduce-rhs}. Part(a) is the distance relation. From the
relation, $f_1$ and $f_2$ are discovered and $f_2$ is trivial.
Parts(b) and (c) are two samples from (a) with the same sample rate
of 1/3. The DDs found from the samples are listed below them. The
trivial DD $f_2$ becomes non-trivial $f_3$ in (b) and non-trivial
$f_4$ in (c).

\begin{table}[h]
\center \caption{Reduction of rhs intervals} \label
{tbl-err-reduce-rhs}
\begin{tabular}{|l|l|l}
 \multicolumn{2}{c}{(a): $\bar{r}$} \\ \hline
 {\bf A} & {\bf B}     \\ \hline
  0 & 2    \\ \hline
  0 & 2   \\ \hline
  0 & 2   \\ \hline
  1 & 2    \\ \hline
  1 & 1   \\ \hline
  1 & 0   \\ \hline
\end{tabular}
\begin{tabular}{l}
 DDs on $\bar{r}$:\\
 $f_1=\aw{A,0}\rar\aw{B,2}$\\
 $f_2=\aw{A,1}\rar\aw{B,{0,2}}$ \\
 $f_2$ is trivial
\end{tabular}
\begin{tabular}{|l|l|l}
 \multicolumn{2}{c}{(b): sample1} \\ \hline
 {\bf A} & {\bf B}     \\ \hline
  0 & 2   \\ \hline
  1 & 0   \\ \hline
  \multicolumn{2}{l}{DDs on sample1:}\\
  \multicolumn{2}{l}{$f_1=\aw{A,0}\rar\aw{B,2}$}\\
  \multicolumn{2}{l}{$f_3=\aw{A,1}\rar\aw{B,0}$}
\end{tabular}
\begin{tabular}{|l|l|l}
 \multicolumn{2}{c}{(c): sample2} \\ \hline
 {\bf A} & {\bf B}     \\ \hline
  0 & 2   \\ \hline
  1 & 1   \\ \hline
  \multicolumn{2}{l}{DDs on sample2:}\\
  \multicolumn{2}{l}{$f_1=\aw{A,0}\rar\aw{B,2}$}\\
  \multicolumn{2}{l}{$f_4=\aw{A,1}\rar\aw{B,1}$}
\end{tabular}
\end{table}

From the example, we can use $f_3$ and $f_4$ to guess $f_2$ by
combining the intervals of the DDs. After the combination, we get
$\hat{f}_2=\aw{A,1}\rar\aw{B,{0,1}}$. This is a way to guessing the
true interval. We define the following operation.

\begin{defi}[DD combination]\label{def:dd-cmb}
Given two intervals $w_1$ and $w_2$, the combination of $w_1$ and
$w_2$, denoted by $w_1\uplus w_2$, is the minimal interval enclosing
both $w_1$ and $w_2$.

Given \emph{sibling} DDs $f_a=\aw{X,w}\rar\aw{A,w_a}$,
$f_b=\aw{X,w}\rar\aw{A,w_b}$, $\ddd$, $f_k=\aw{X,w}\rar\aw{A,w_k}$,
the combination of the DDs, denoted by $f_a\uplus f_b\uplus \ddd
\uplus f_k$, is defined to be $\aw{X,w}\rar\aw{A,w_a\uplus w_b\uplus
\ddd \uplus w_k}$.\done
\end{defi}

Following the combination operation, as the number of samples
increases, the interval of $\hat{f}_2$ becomes closer to that of
$f_2$. We now analyze how the number of samples affect the errors in
guessing the right interval.

A tuple is allowed to be included in a sample only once and the
chance for this to happen is the sample rate $\varrho={N_s\over N}$
where $N_s$ is the sample size which is the same for all samples. For
each sample, a tuple is either in or out. A tuple may be included in
many samples. Thus, for a tuple to be included in $ks$ of $ns$
samples follows the binomial distribution $B(ks)=\combin{ns}{
ks}\varrho^{ks}(1-\varrho)^{ns-ks}$. The probability for a tuple to
be included in one or more samples is $B(ks\ge1)=1-\combin{ns}{
0}\varrho^{0}(1-\varrho)^{ns-0}$=$1-(1-\varrho)^{ns}$. From this we
have
 \[n_x={ln(1-B) \over ln(1-\varrho)}\]
where $B$ is short handed notation for $B(ks\ge1)$. If $\varrho=10\%$
and $B=90\%$, $ns$ needs to be $22$. If we reduce the sample rate to
$\varrho=1\%$, $ns$ must be $230$.

This formula indicates that when the computation is possible, a large
sample rate should be used to reduce the number of samples needed.

When the DDs from multiple samples are combined, the probabilities of
missed and unwanted DDs will change. The chance for a wanted DD $f$
to be missed from a sample is $P_{m}=P(N,M_f,N_s, k_f< \theta N_s)$
by Formula (\ref{eq:missed wanted}). The chance of missing $f$ in all
the $ns$ samples is $P_{mw}^{ns}$. It is easy to see that
$P_{m}>P_{m}^{ns}$ if $ns>1$. The more samples are used, the less
possible that $f$ is missed.

Similarly, by Formula (\ref{eq:found-unwanted }), the chance of
finding an unwanted DD $f$ in all samples is $P_{uw}=1-P_{m}^{ns}$.
As the number $ns$ of samples increases, the probability $P_{uw}$ for
$f$ to be discovered gets larger.

After DDs are discovered from samples and are combined, we use three
filters to remove the ones that are possibly erroneous. These filters
are (1) the support, (2) the ratio of the number of the samples from
which a DD is found over the total number of samples, and (3) the
ratio of the average interval width in samples over the combined
interval width of all samples. Our experiments show that the second
filter can best minimize the errors and next one is the third filter
and the support does not work well. The details will be shown in the
experiments section.

The above sampling analysis is based on the distance table $\bar{r}$.
Computing all tuples of $\bar{r}$ from the raw data $r$ is too
expensive and we need to avoid this. We recall the way in which
$\bar{r}$ is computed. Assume that each tuple of $r$ has an index
starting from 0, the tuples are ordered by the index, and $n=|r|$.
The first tuple of $r$ is paired with all its following tuples to
generate the first $(n-1)$ tuples of $\bar{r}$. Then the second tuple
of $r$ is paired with its following tuples of $r$ to compute the next
$(n-2)$ tuples of $\bar{r}$. After the $x$-th tuple of $r$ is paired
with its following tuples, the total number $i$ of tuples generated
is equivalent to the area of a quadrilateral with a pair of parallel
edges and two neighboring right angles. Thus $i=[(n-1)+(n-x-1)]*
x/2$. To determine the $x$-th and $y$-th tuples of $r$ for the $i$-th
tuple of $\bar{r}$, we use
\begin{align}
 \begin{aligned}\label{eq:index-r}
 & x=int({(2n-1)-\sqrt{(2n-1)^2-8*i}\over 2})\\
 & di=i-{(2n-1-x)*x \over 2} \\
 & if\ (di>0): \ \ y=x+di; \\
 & else:\ \{\ x=x-1; \ \ y=(n-1);\ \}
 \end{aligned}
\end{align}
where $int()$ is the truncate function, $di\ge 0$, the indexes $x$
and $y$ start from 0, and the index $i$ starts from 1. Using the formula,
only the distances of tuple pairs that are drawn to be in the sample
need to be computed.

\section{Experiments}\label{sec:experiments}
Our experiments are done on a laptop computer with Intel i5-2520M
CPU@2.5GHz, 8GB of main memory, and Windows 7 OS. The programming
language used in the implementation is Java with JDK 1.7.

We use the experiments to demonstrate the following points.
\begin{list1}
  \item How our algorithm performs to the change to the number of
  tuples and the number of attributes in data.
  \item How the parameters, support, satisfaction threshold $\epsilon$
  and the interestingness,
  affect the results and efficiency performance.
  \item How sampling affects the DDs found in comparison to
  non-sampling cases.
\end{list1}

\subsection{Data sets and distance functions}\label{subsec:data}
We use three data sets, the DBLP, the US Census data, and the US
Airline data. (1) The DBLP data is the publication reference data in
XML obtained from http://dblp.uni-trier.de/xml/. This data is then
transformed to the relational format having 6 columns namely the type
(journal or conference), number of authors, author names, title,
journal/conference name, and year. The volume of data is huge. We
downloaded a portion of the data containing 40,000 tuples. (2) The
Adult data is the US Census data obtained from
http://mlr.cs.umass.edu/ml/machine-learning-databases/adult/ which
has 15 columns and 32,000 tuples. (3) The US Airline data is about
the airline on-time running statistics. The data was obtained from
http://www.transtats.bts.gov. After the columns containing large
amount of null values and repeating values are removed, the data has
20 columns covering dates, airlines and flights, departure and
destination airports, and on-time running information. The data has
about 500,000 tuples.

The distance functions $d_X(t_1,t_2)$ for the attributes of the data
sets are important to the discovery. Different differential functions
lead to different computation sizes, different DDs discovered, and
different amount of time used in the discovery. Table
\ref{table-n-intervals} lists the distance functions used for the
attributes of the data sets, and the number of intervals (nIntv) for
the attributes. In the table, `wordDif' means that the distance is
the number of different words in the two tuples for the attribute,
e.g., the distance between 'Hello World' and 'Hello Helen' is 2.
`numeDif/5' means that the distance is the difference of numeric
values of two tuples for the attribute divided by 5. `hierDis'
indicates that the distance is the number of edges of the shortest
path between the two nodes in the taxonomy hierarchy of the
attribute.If the taxonomy for WorkClass is
$WorkClass(neverWorked)(worked (withPay)(withoutPay))$ where the
notation $A(B)$ represents that $B$ is a child of $A$, the distance
between $neverWorked$ and $withPay$ is 3. `Bool' indicates that the
distance is 0 (same) or 1 (different).

We understand that there are many possible ways to define the
distance functions. The domain expertise and computation capabilities
need to be taken into consideration when the functions are designed.

\begin{table}
\caption{Number of intervals of data sets}\label{table-n-intervals}
\begin{tabular}{|l|l|l|}
  \multicolumn{3}{l}{DBLP}\\\hline
  {\bf Attr} & {\bf $d_A(t_1,t_2)$} & {\bf nIntv} \\\hline
  Type & wordDif & 2 \\\hline
  Title & wordDif & 31 \\\hline
  NAuth & numeDif & 11 \\\hline
  Authors & wordDif & 23 \\\hline
  Forum & wordDif & 18 \\\hline
  Year & numeric & 43 \\\hline
\end{tabular}
\begin{tabular}{|l|l|l|}
  \multicolumn{3}{l}{Adult}\\\hline
  {\bf Attr} & {\bf $d_A(t_1,t_2)$} & {\bf nIntv} \\\hline
  Age & numeDif/5 & 14 \\\hline
  WorkClass & hierDist & 6 \\\hline
  Wedge & numeDif/1000 & 101 \\\hline
  Education & hierDist & 15 \\\hline
  EducNumb & numeDif & 15 \\\hline
  Marital & wordDif & 3 \\\hline
  Occupa & wordDif & 2 \\\hline
  ... & ... & ... \\\hline
\end{tabular}
\begin{tabular}{|l|l|l|}
  \multicolumn{3}{l}{USair}\\\hline
  {\bf Attr} & {\bf $d_A(t_1,t_2)$} & {\bf nIntv} \\\hline
  FlightNumb & Bool & 31 \\\hline
  DepTime & numeDif & 12 \\\hline
  DepDelay & numeDif/15 & 47 \\\hline
  Distance & numeDif/100 & 32 \\\hline
  ArrTime & numeDif & 12 \\\hline
  ... & ... & ... \\\hline
\end{tabular}
\end{table}

\subsection{Time performances}
We implemented our algorithm called the {\bf Lattice algorithm} and
also the algorithm proposed in \cite{diffDep-tods11} called the {\bf
Split algorithm} (we give it such a name because it splits the search
space). We note that the Lattice and the Split algorithms find
different DDs as shown in detail in the related work section. The
comparison of the time performances of the two algorithms is not very
meaningful. The reason for including the Split algorithm in the experiments
is to show our respect to existing work.

\begin{figure}
  \center
  \includegraphics[scale=.9]{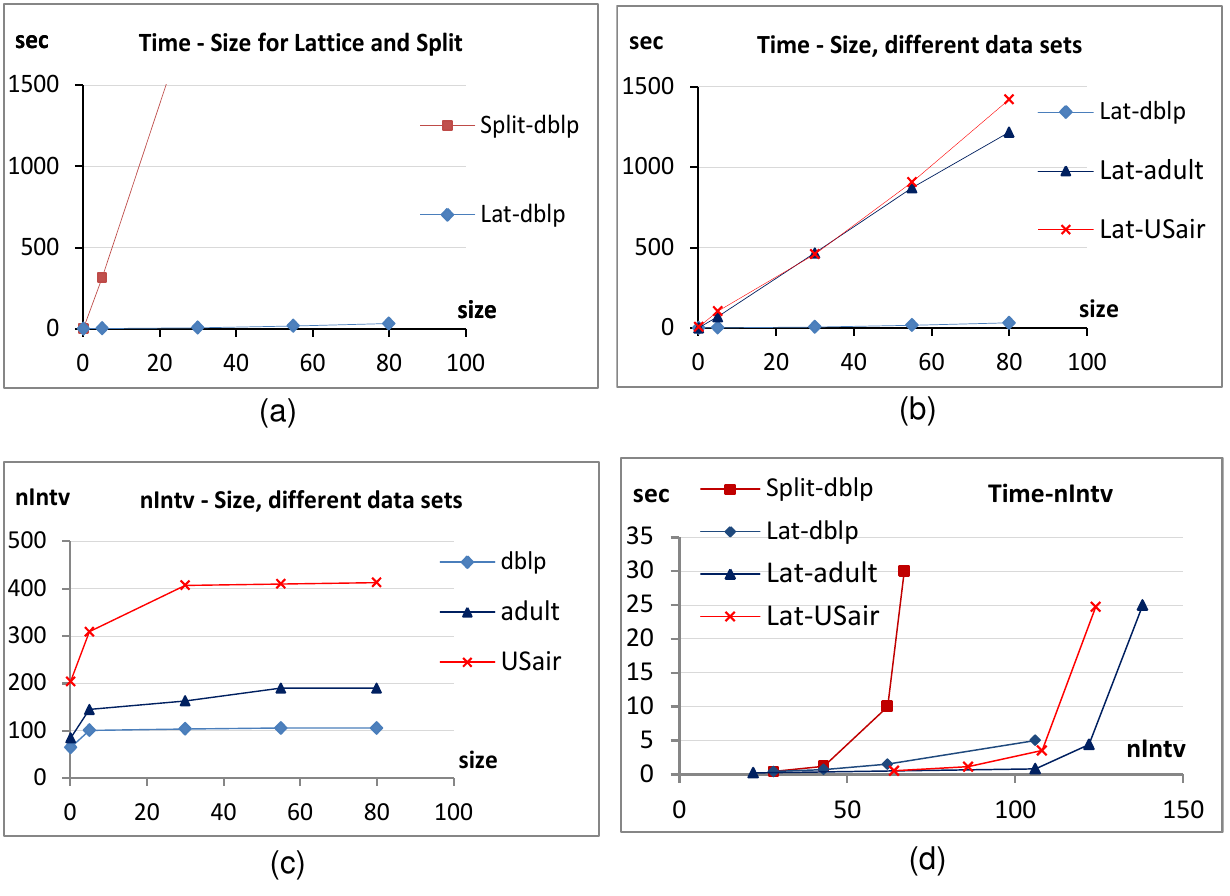}
  \caption{Time to data sizes \label{fig:time-size}}
\end{figure}

Figure \ref{fig:time-size} shows the experiment results. Figure
\ref{fig:time-size}(a) and (b) are about the time (in seconds) used
for DD discovery versus the data size in the number of tuple pairs of
the data divided by 1000. Part(a) shows the time used by the two
algorithms on the DBLP data. The line `Split-dblp' for the Split
algorithm is very sensitive to data size (and even more to the number
of attributes, as shown in (d)), and used more time and ran out of
memory very soon. Because of this, it was not tested against other
data sets. Part(b) is the time performance of our Lattice algorithm
on DBLP, Adult, and USair data. Because the low number of attributes
of the DBLP data, the line `Lat-Dblp' is almost flat. For the other
two data sets, the lines are close to straight.

We notice that the data size change causes another complication. As
the data size increases, the number of attribute intervals also
increases (Figure \ref{fig:time-size}(c)). The increased number of
intervals has impact on the time performance. This means that the
time-size performance in (a) and (b) also includes the effect caused
by some increase of the number of intervals. The number of intervals
gets stable when the size reaches about `30'.

The relationship between the time and the number of attributes is
shown in Figure \ref{fig:time-size}(d). The data size was fixed to
`80' in the experiments for the chart. The x-axis is labeled with
`nIntv' which means the sum of the number of intervals of all
attributes. The reason for showing the number of intervals instead of
the number of attributes is that the former has the actual time
performance impact. An attribute with 40 intervals has much more
impact on the performance than 5 attributes each with only 2
intervals. In the chart, when the number of intervals gets to more
than 100, the performance gets worse very quickly for the Lattice
algorithm. We note that the `Lat-dblp' line is shorter because the
data set has less total number of intervals for 6 attributes than
other data sets. At the same time, when the total number of intervals
exceeds 70, `Split-dblp' ran out of memory.

\subsection{The effect of the parameters}
The implementation uses three parameters, namely the number of
interesting intervals, the support, and the satisfaction threshold as
input to experiment cases for our Lattice algorithm.

Setting a threshold for the number of interesting intervals for an
attribute means to reduce the total number of intervals in the
calculation. The effect of this is already shown in Figure
\ref{fig:time-size}(c).

\begin{figure}
  \center
  \includegraphics[scale=.9]{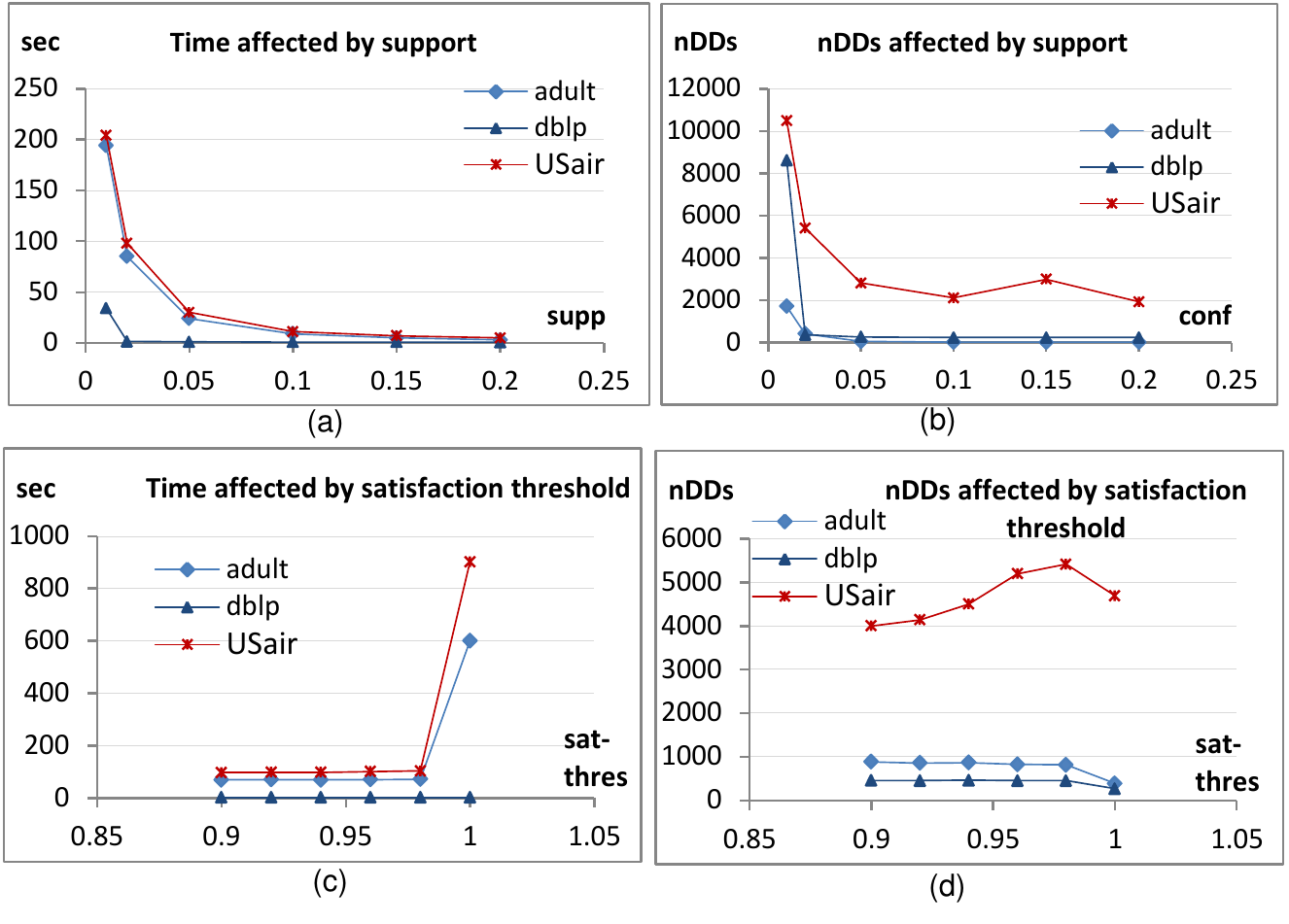}
  \caption{Time affected by support \label{fig:time-supp}}
\end{figure}

As in data mining, the effect of support has significant impact on
performance. Setting a minimal support threshold can change many
non-computable cases to computable. We did an analysis using our
Lattice algorithm and the result is shown in Figure
\ref{fig:time-supp} (a) and (b) where `nDDs' means the number of DDs
discovered from data. From Figure \ref{fig:time-size}(c), we know
that the total number of intervals for the USair data is much larger.
This leads to much more DDs found from the data too.

Setting an approximate satisfaction threshold (less than but close to
100\%) has two effects. One is that the number of non-trivial DDs in the
output is increased. The reason is that some DDs are trivial in the
case of full satisfaction and become non-trivial in the case of
approximate satisfaction. The other effect is that the time used to
find approximately satisfied DDs is less than that for fully
satisfied DDs. The reason behind this is that more discovered DDs
cause more pruning. These two effects are shown in Figure
\ref{fig:time-supp}(d) and (c) respectively. The axis `sat-thres'
means satisfaction threshold and when it is 1, the satisfaction is
full.

\subsubsection{Data quality problems} We run experiments on the adult
data with two values for the approximate satisfaction threshold:
$\epsilon=1$ (full satisfaction) and $\epsilon=0.95$ (approximate
satisfaction) while all other parameters, including the support, are
fixed. We then compare the results. For each fully satisfied DD
$f_1=\aw{X,w}\rar\aw{A,w_1}$ and its approximately satisfied partner
DD $f_2=\aw{X,w}\rar\aw{A,w_2}$, we calculate the ratio $ratio={w_2
\over w_1}$ which indicates the amount left after the shrink. This
ratio is then used to rank all fully satisfied DDs. DDs with low
ratios mean that by excluding a very small portion of tuple pairs
(5\%), the rhs intervals of the DDs become much smaller.

Table \ref{table-interval-shrink} gives some examples of the DDs. The
first column of the table is the support of the fully satisfied DDs,
the second column is the ratio, and the third column is $f_2$. We
choose to explain the first two DDs. The first DD in the table says
that when satisfaction threshold is reduced from 1 to .95, the
interval of $eduNum$ is reduced by 50\%. That is, 5\% of tuple pairs
spreads their $eduNum$ values in [6,12]. We see that these 5\% of
tuples are outliers. After excluding the outliers, most people who
work for the same number of hours have an education difference no
more than 6. This rule is supported by 26\% of total tuple pairs.

The second one says that for 95\% of people pairs. if they make the
same amount of capital loss, they make the amount of gain. Only 5\%
of people (outliers) are different. This rule is fully supported by
all pairs.

\begin{table}
\caption{Interval shrink when approximate threshold is set to
.95.}\label{table-interval-shrink}
\begin{tabular}{|l|l|l|}  \hline
 {\bf support} &  {\bf ratio} &  {\bf approximate DD} \\ \hline
 26.52  & 50.00 & $\aw{hrsWk,0}\rar \aw{eduNum,{0,6}}$ \\ \hline
 100.00 & 33.33 & $\aw{capLoss,0}\rar \aw{capGain,0}$ \\ \hline
 18.52  & 75.00 & $\aw{relat,0}\aw{race,0}\rar \aw{marital,{0,2}}$ \\ \hline
  9.79  & 42.85 & $\aw{occup,0}\rar \aw{eduNum,{0,5}}$\\ \hline
\end{tabular}
\end{table}

\subsection{Errors of sampling}
Sampling enables the discovery problems that are impossible to
compute to become computable and plays an important role in data mining and
data analysis. The aim of experiments here is to compare the DDs
discovered from the whole data with the DDs found from the samples to
analyze the errors caused by sampling.

Given a relation $r$, a sample size $N_s$, and the number of samples
$ns$, following Formula (\ref{eq:index-r}) repeatedly, we draw $ns$
samples, $s_1, \ddd, s_{ns}$, from $r$. Let $D^{\bar{r}}_\theta$
denote the set of DDs found from the distance relation $\bar{r}$ of
$r$ directly with the support threshold $\theta$, and
$D^{s_i}_\theta$ the set of DDs discovered from the sample $s_i$. Let
$D^{s}_\theta$ denote the set of DDs combined from the DDs discovered
from all the samples following Definition \ref{def:dd-cmb}.

The number of missed DDs is $|D^{\bar{r}}_\theta- D^{s}_\theta|$. The
relative error rate of missed DDs is
\begin{align}\label{eq:err-rate-w}
    \begin{aligned}
      err_{m}={|D^{\bar{r}}_\theta- D^{s}_\theta| \over
      |D^{\bar{r}}_\theta|}
    \end{aligned}
\end{align}

Similarly, the number of unwanted DDs is $|D^{s}_\theta -
D^{\bar{r}}_\theta|$ and the error rate is
\begin{align}\label{eq:err-rate-u}
    \begin{aligned}
      err_{uw}={|D^{s}_\theta- D^{\bar{r}}_\theta| \over
      |D^{\bar{r}}_\theta|}
    \end{aligned}
\end{align}

We now present the experiments we conducted to analyze some
properties of sampling with the DBLP data.

\begin{figure}
  \center
  \includegraphics[scale=.9]{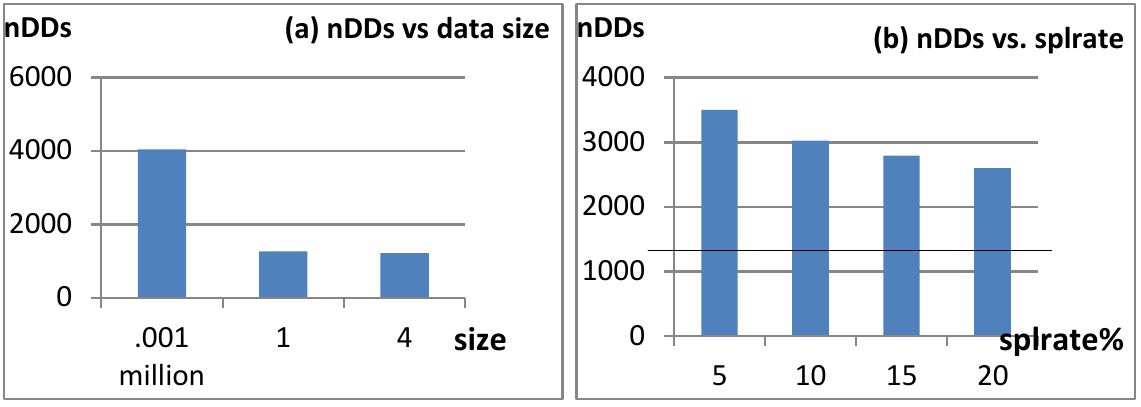}
  \caption{The number of DDs discovered \label{fig:nDDs-splrate}}
\end{figure}

The first experiment analyzes how the data sizes and sample rates
affect the number of DDs discovered and the results are shown in
Figure \ref{fig:nDDs-splrate}. Part (a) of the figure shows that as
data size increases, with the support threshold $\theta=0.5\%$, the
number of DDs discovered from the unsampled data,
$D^{\bar{r}}_\theta$, decreases. Part (b) shows that as the sample
rate increases, the number of DDs discovered from the samples,
$D^{\bar{r}}_\theta$,  also decreases. We note that if zero support
threshold is used, the number of DDs discovered from unsampled data
and the samples will increases because the number of possible
intervals will increase.

\begin{figure}
  \center
  \includegraphics[scale=.9]{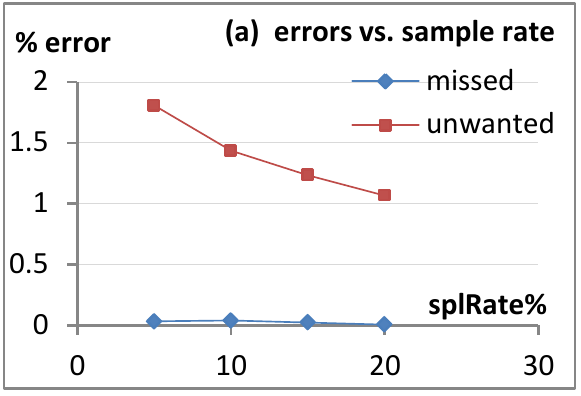}
  \caption{The errors of sampling \label{fig:sample-errs}}
\end{figure}

The second experiment is about errors of sampling rate and the
results are shown in Figure \ref{fig:sample-errs}. It shows how the
error rates (Formulas (\ref{eq:err-rate-w}, \ref{eq:err-rate-u}))
change with the sample rate. It shows that the error rate of missed
wanted DDs is much lower than the found unwanted error rate. As we
explained before, the number of unwanted DDs is much more than the
number of missed DDs. The figure also shows that as the sample rate
increases, the overall error rates (both missed and unwanted) reduce
too. Readers may feel confused as we said in Formulas (\ref{eq:missed
wanted} and \ref{eq:found-unwanted }) that the errors of missed
wanted and found unwanted are complementary. We note that
complementary formulas are about the error possibilities of
individual DDs. What we present in this diagram is about the number
of all DDs. We believe that the down trend of errors with the
increasing sample rate is caused by the fact that the number of DDs
decreases as the sample rate gets larger and that the errors in DDs
are reduced even faster. This is the evidence for our conclusion that
large sample size should be used if the computation is practical.

\begin{figure}
  \center
  \includegraphics[scale=.9]{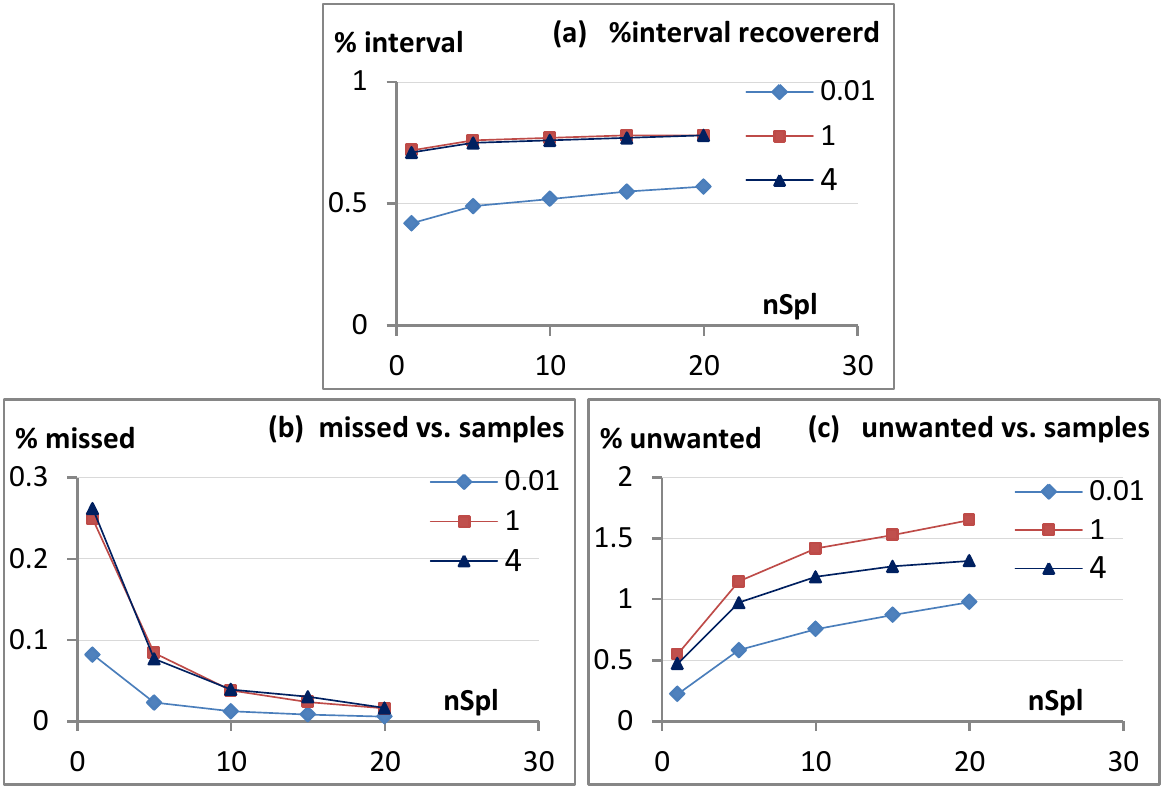}
  \caption{The errors for different number of samples \label{fig:nspl-err}}
\end{figure}

A further experiment is on the relationship between errors and the
number of samples and this is shown in Figure \ref{fig:nspl-err}. This
experiment is done with fixed sample rate, fixed support and fixed
data size.  Parts (a) and (b) show that as the number of samples
increases, the intervals of attributes are recovered better and the
errors of missed wanted DDs get smaller. However, Part (c) shows that
the increasing number of samples causes the errors of found unwanted
DDs to grow.

\begin{figure}
  \center
  \includegraphics[scale=.9]{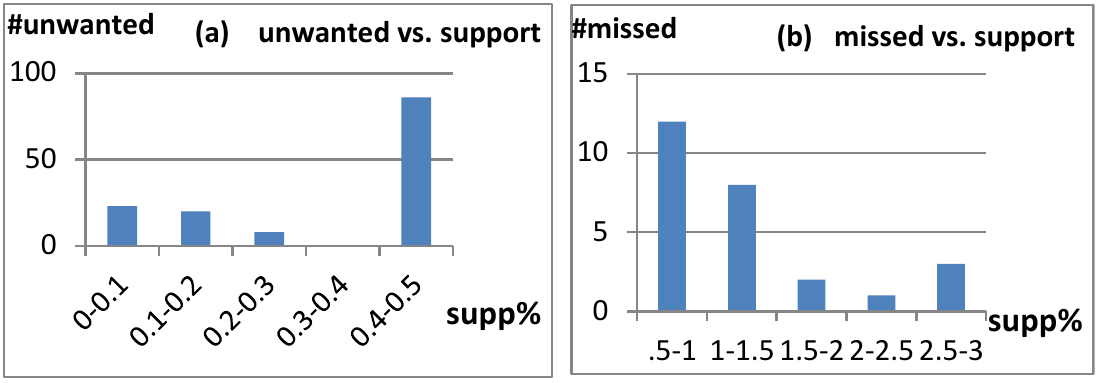}
  \caption{The relationship between errors and support \label{fig:supp-err}}
\end{figure}

Another experiment analyzes the relationship between the errors and
the support and the result is shown in Figure \ref{fig:supp-err}. The
experiment is done with the support threshold of 0.5\%. The errors in
terms of missed wanted and found unwanted DDs are derived and then
these erroneous DDs are grouped by these support. Part(a) shows that
although any unwanted DDs with the support between 0 and 0.5\% can be
found as errors, most of errors are from the DDs whose support is
close to the threshold. In the same way, among the missed wanted DDs,
the number with close to threshold support is more than the number
with higher support.

\begin{figure}
  \center
  \includegraphics[scale=.9]{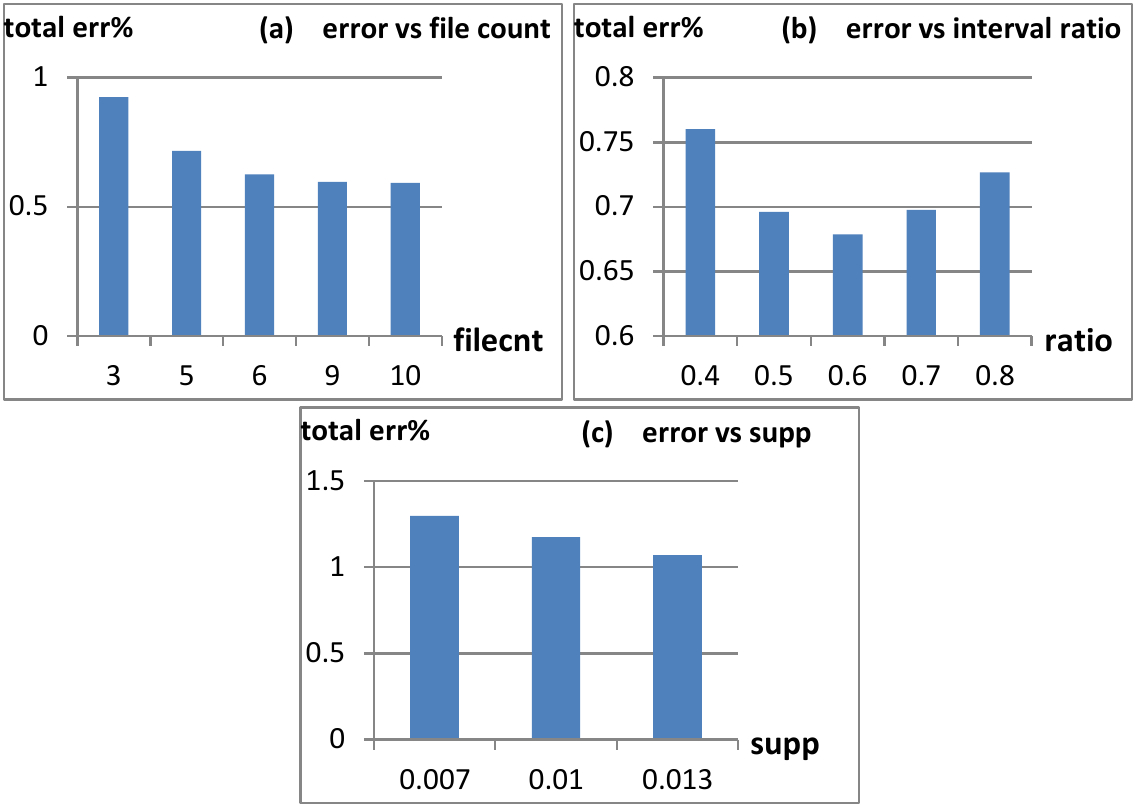}
  \caption{Filtering insignificant DDs \label{fig:filtering-errs}}
\end{figure}

The final experiment investigates whether there is a possible way to
filter some of the error DDs from the discovery. For a set of
discovered DDs $D^{s}_\epsilon$ from 10 samples, we use three types
of filters. The first type is the count of the files from which a DD
is found. If a DD is found from one of the samples but not from
others, this DD may not be significant. The result of this way of
filtering is shown in Figure \ref{fig:filtering-errs}(a). The
vertical axis is the relative error rate of the number of missed and
the number of unwanted. The horizontal axis labeled by 'filecnt' the
the file count used in filtering. When the file count gets to 6 or
more (out of 10 samples), the total error rate reduces to the minimal.

The second way of filtering uses an interval ratio and the result is
in Figure \ref{fig:filtering-errs}(b). Suppose that for DD
$\aw{X,w}\rar \aw{A,w'}$, we find $w'=[1,2]$ from the first sample
and $w'=[3,4]$ from the second sample. Then the average width of $w'$
for each sample is 2 and the combined interval is 4. The interval
ratio then is 2/4. If the ratio is small over many samples, each
sample only get a small portion of a large interval and the DD is
less significant. Figure \ref{fig:filtering-errs}(b) indicates that
this way of filtering achieves the minimal error rate when when the
ratio is 0.6. A larger ratio will cause the loss of more missed DDs
and this is undesirable.

The third way of filtering is the most direct way which uses support
of the discovered DDs. A filtering threshold (like 0.007) is used to
filter away the DDs if their average support over all samples is less
than the filtering threshold. The result in Figure
\ref{fig:filtering-errs}(b) shows that this way of filtering is not
as good as the other two because if a larger filtering threshold is
used, too few DDs remain.

\section{Related work}\label{sec:related}

On the discovery of functional dependencies (FDs), many algorithms
have been developed for this purpose. The major ones are TANE
\cite{huht99cj-tane-FD-AFD}, FD\_Mine \cite{yao08dmkd-minFD}, FUN
\cite{nov01icdt-discv-EmbFD}, hash-based \cite{FDdiscHash-dke13liu},
Dep-Miner \cite{lopes00edbt-eff-discvFD}, and FastFDs
\cite{wyss01dawak-fastFDs} etc.. FDs hold at the schema level and the
algorithms for discovering FDs do not apply to DDs.

In recent years, a few new types of dependencies, XML functional
dependencies, conditional functional dependencies, matching
dependencies, and differential dependencies, have been proposed for
all sorts of purposes. Some discovery algorithms are also proposed
for these types of dependencies:
\cite{yuC08vldbj-discoverXFD,XCFD-d-adc11} for XML functional
dependencies,
\cite{fan10tkde-CFDd,yeh10vldb-cfdDiscv,jiuyong12-cfd-disc,XCFD-d-adc11}
for conditional functional dependencies,
\cite{MatchDepDisc-cikm09song,MatchDepDisc-arXiv09song} for matching
dependencies, and \cite{diffDep-tods11} for differential
dependencies.

One work that relates to ours is the algorithm discovering matching
dependencies in \cite{MatchDepDisc-arXiv09song}. We note that
matching dependencies are special cases of differential dependencies,
which is what our work aims to discover. The algorithm in
\cite{MatchDepDisc-arXiv09song} transforms the raw data into a
distance data set and then discovers a matching dependency for a {\bf
given} $X\rar Y$ based on the relative frequencies of distance
tuples. It searches for the thresholds (interval boundaries in our
terms) for $X$. The assumption with a given candidate restricts the
general problem with the search space size $2^{|R|*d}$ to a very
specific problem with the search space size $d^{|X|}$ where $d$ is
the average number of possible intervals among all attributes of
schema $R$. Our work does not assume any given $X\rar Y$ and
therefore is much more general.


The algorithm in \cite{diffDep-tods11} on DD discovery is the most
related work to what we do in this paper. Its core reduction
algorithm works by fixing the rhs attribute and its interval of a
candidate DD, and then splitting a given search space of the left
hand side to find lhs while pruning the subspaces that are not
possible to contain any lhs. The algorithm assumes ``that
differential functions in $\Phi(X)$ has already been arranged by
subsumption order". We see that the size of the search space is
large. If only the intervals containing 0 distance are considered,
and each attribute has $m$ such intervals, there are $m^{|X|}$ number
of points in the search space. This space size becomes larger when
the intervals are defined in other ways.

\begin{figure}\label{fig:diff}
  \center
  \includegraphics[scale=.9]{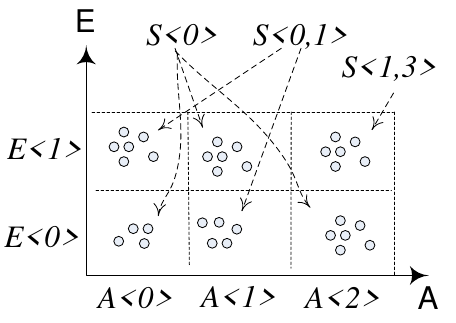}
  \caption{Difference}
\end{figure}

In addition to the method and the representation differences, the two
methods find different DDs. For the distance data shown in Figure
\ref{fig:diff} where $E$ stands for $Edu$, $A$ $Age$, and $S$ $Sal$.
A cell labeled with $\aw{S,{0,1}}$ means that the distances of the
tuple pairs falling in the cell are between 0 and 1. With this data,
our method finds the following DDs.
 \begin{tab}
 $\aw{Age,0}\aw{Edu,0}\rar \aw{Sal,0}$\\
 $\aw{Age,1}\aw{Edu,1}\rar \aw{Sal,0}$\\
 $\aw{Age,2}\aw{Edu,0}\rar \aw{Sal,0}$\\
 $\aw{Age,0}\aw{Edu,1}\rar \aw{Sal,{0,1}}$\\
 $\aw{Age,1}\aw{Edu,0}\rar \aw{Sal,{0,1}}$\\
 $\aw{Age,2}\aw{Edu,1}\rar \aw{Sal,{1,3}}$
 \end{tab} The
Split method find the DDs of
 \begin{tab}
 $\aw{Age,0}\aw{Edu,0}\rar \aw{Sal,0}$\\
 $\aw{Age,{0,2}}\aw{Edu,0}\rar \aw{Sal,{0,1}}$\\
 $\aw{Age,{0,1}}\aw{Edu,{0,1}}\rar \aw{Sal,{0,1}}$\\
 $\aw{Age,{0,2}}\aw{Edu,{0,1}}\rar \aw{Sal,{0,3}}$
 \end{tab}
Both sets of DDs are correct with regard to the DD definition and the
respective methods. The Split method finds less DDs but our method
identifies dense groups better.

The Split method may be used with our base intervals. In that case,
if $Sal$ has $m$ base intervals, the number of rhs for $Sal$ to be
considered is the factorial of $m$ (i.e., $m!$). In contrast to the
case where the intervals starting with 0 are used, only $m$ rhs
intervals need to be considered. As the performance for
starting-with-0 intervals is already very poor, adding the complexity
makes it simply not computable.

\section{Conclusion}\label{sec:conclusion}
This paper proposes an algorithm for discovering differential
dependencies from data. The algorithm works based on traversing of a
lattice level by level and uses a number of way to prune implied DDs
to reduce the computation size. The lattice is different from the
lattice used in association rule and functional dependency discovery
in that edges in the lattice do not represent candidate differential
dependencies.

We also conducted a comprehensive analysis on the errors caused by
sampling in the discovery computation and propose ways to filter out
possible errors.

The proposed algorithm still have high complexity. The future work of
this paper includes the investigation of further pruning methods so
that uninteresting DDs can be avoided. Work is also needed to
investigate how DDs can be used to repair low quality data values and
to identify schema mapping and related relations.

\bibliography{DD-sati.bbl}
\bibliographystyle{plain}

\end{document}